\documentclass[10pt,aps,prd,twocolumn,showpacs,amsmath,amssymb,nofootinbib,eqsecnum,preprintnumbers]{revtex4-1}

\usepackage[utf8]{inputenc}				

\usepackage{xcolor} 					

\usepackage{mathtools}					
\usepackage{tensor}						
\usepackage{suffix} 					
\usepackage{xargs} 						
\usepackage{relsize} 					
\usepackage[mathscr]{eucal} 			

\newcommand\bs[1]{\boldsymbol{#1}}

\newcommand\dd{\mathrm{d}}
\newcommand\pp{\partial}

\renewcommandx{\t}[3][1={}, 3={}]{\tensor[#1]{#2}{^{\mathstrut}_{\mathstrut}#3}}
\newcommandx{\tb}[3][1={}, 3={}]{\t[#1]{}\mkern-1mu\bs{#2}\mkern-1mu\t{}[#3]}

\WithSuffix\newcommandx\t*[3][1={}, 3={}]{\tensor*[#1]{#2}{^{\mathstrut}_{\mathstrut}#3}}
\WithSuffix\newcommandx\tb*[3][1={}, 3={}]{\t*[#1]{}\mkern-1mu\bs{#2}\mkern-1mu\t*{}[#3]}

\newcommand{\pow}[2]{{#1}^{#2}{}\!\!\mkern-1mu}

\newcommand{\nablai}{\nabla\!\mkern-1mu}

\newcommand{\nablan}[1]{{\nabla}^{#1}{}\!\!\!\mkern-1mu}

\newcommand{\con}{\mkern-1mu\cdot\mkern-1mu}

\newcommand{\ind}[1]{\mathsmaller{#1}\mkern-1mu}

\newcommand{\ti}[1]{{\mathstrut}^{\mathstrut}\smash{\t[^{#1}]{}\!}}

\newcommandx{\tvec}[2][2={}]{\frac{\smash{\tb{\pp}[#2]}}{\bs{\pp}#1}}

\newcommandx{\ev}[2][2={}]{\t*{\bs{e}}[^{#2}_{#1}]}
\newcommandx{\ef}[2][2={}]{\t*{\bs{e}}[^{#1}_{#2}]}
\newcommandx{\epsv}[2][2={}]{\t*{\bs{\epsilon}}[^{#2}_{#1}]}
\newcommandx{\epsf}[2][2={}]{\t*{\bs{\epsilon}}[^{#1}_{#2}]}

\newcommandx{\evh}[2][2={}]{\t*{\bs{\hat{e}}}[^{#2}_{#1}]}
\newcommandx{\efh}[2][2={}]{\t*{\bs{\hat{e}}}[^{#1}_{#2}]}
\newcommandx{\epsvh}[2][2={}]{\t*{\bs{\hat\epsilon}}[^{#2}_{#1}]}
\newcommandx{\epsfh}[2][2={}]{\t*{\bs{\hat\epsilon}}[^{#1}_{#2}]}

\newcommand{\Ric}{\mathrm{Ric}}

\newcommand\lie[2]{\pounds\smash{\raisebox{-1ex}{$#1$}}#2}

\newcommand\imag{i}

\newcommand\op[1]{\mathsf{#1}}

\newcommand\gden{\mathfrak{g}}

\newcommand\switch[2]{(#1\leftrightarrow #2)}

\newcommand\pois[2]{\{#1\,{,}\,#2\}}
\newcommandx\sn[3][3={}]{[#1\,{,}\,#2] \t*{}[_{\mathsmaller{\mathrm{SN}}}^{#3}]}
\newcommand\com[2]{[ #1\,{,}\,#2 ]}

\newcommand\feq{\mathrel{\phantom{=}}}

\newcommand{\shiftandleft}[1]{\hspace{#1}&\hspace{-#1}}

\newcounter{notenumber}

\begin{document}

\title{Weak electromagnetic field admitting integrability in Kerr--NUT--(A)dS spacetimes}

\author{Ivan Kol\'a\v{r}}
\email{ivan.kolar@utf.mff.cuni.cz}
\affiliation{Institute of Theoretical Physics,
Faculty of Mathematics and Physics, Charles University in Prague,
V~Hole\v{s}ovi\v{c}k\'ach 2, 180 00 Prague, Czech Republic}

\author{Pavel Krtou\v{s}}
\email{Pavel.Krtous@utf.mff.cuni.cz}
\affiliation{Institute of Theoretical Physics,
Faculty of Mathematics and Physics, Charles University in Prague,
V~Hole\v{s}ovi\v{c}k\'ach 2, 180 00 Prague, Czech Republic}

\date{May 21, 2015} 

\begin{abstract}
We investigate properties of higher-dimensional generally rotating black-hole spacetimes, so-called Kerr--NUT--(anti)-de Sitter spacetimes, as well as a family of related spaces which share the same explicit and hidden symmetries. In these spaces, we study a particle motion in the presence of a weak electromagnetic field and compare it with its operator analogies. First, we find general commutativity conditions for classical observables and for their operator counterparts, then we investigate a fulfillment of these conditions in the Kerr--NUT--(anti)-de Sitter and related spaces. We find the most general form of the weak electromagnetic field compatible with the complete integrability of the particle motion and the comutativity of the field operators. For such a field we solve the charged Hamilton--Jacobi and Klein--Gordon equations by separation of variables.
\end{abstract}


\maketitle

\section{Introduction}\label{sc:intro}

Investigation of higher-dimensional black-hole spacetimes has become important not only in connection with the string theory \cite{EmparanReall:2008}, but also due to the fact that they serve as nontrivial examples of integrable systems.

The most general known geometries describing higher-dimensional, generally rotating black holes with spherical horizon topology, Newman--Unti--Tamburino (NUT) parameters and an arbitrary cosmological constant are Kerr--NUT--(anti)-de Sitter [(A)dS] spacetimes \cite{ChenLuPope:2006}. Such spacetimes have many remarkable properties. Namely, they possess a nondegenerate closed conformal rank-two Killing--Yano tensor (the so-called principal Killing--Yano tensor) \cite{KubiznakFrolov:2007}. It was shown that this object generates the tower of explicit and hidden symmetries\footnote{The name \textit{hidden symmetries} is used in the literature in various contexts. In this paper we mean by hidden symmetry the existence of a conserved quantity for the geodesic motion which is not linear in momenta. If such a quantity is a simple power of the momentum, it is generated by a Killing tensor. By hidden symmetry we thus mean the existence of a Killing tensor of rank two (or higher). The explicit symmetry is generated by a Killing vector (i.e., by a Killing tensor of rank one) and, contrary to hidden symmetries, it corresponds to a spacetime isomorphism.}
associated with Killing vectors and rank-two Killing tensors \cite{KrtousEtal:2007a}. This exceptional symmetry of \mbox{Kerr--NUT--(A)dS} enables integrability of geodesic motion \cite{PageEtal:2007} as well as separability of the Hamilton--Jacobi, Klein--Gordon \cite{FrolovEtal:2007,SergyeyevKrtous:2008}, and Dirac equations \cite{CarigliaEtal:2011b}.

It turns out that the structure of spacetime symmetries following from the existence of the principal Killing--Yano tensor is rather restrictive. It was proved in \cite{HouriEtal:2008,HouriEtal:2007,KrtousFrolovKubiznak:2008} that any spacetime admitting the principal Killing--Yano tensor can be written in the off-shell \mbox{Kerr--NUT--(A)dS} form. In this paper, we demonstrate that there exist geometries which have the same explicit and hidden symmetries as the \mbox{Kerr--NUT--(A)dS} spacetimes, but without having the principal Killing--Yano tensor.

Since the higher-dimensional generalization of the charged black-hole spacetime has not been found yet, some effort was devoted to the study of weakly charged higher-dimensional black holes, in particular \mbox{Kerr--NUT--(A)dS} spacetimes with a test electromagnetic field \cite{AlievFrolov:2004,Aliev:2006a,Aliev:2006, Aliev:2007,Krtous:2007}. Although this electromagnetic field does not affect the geometry, it can dramatically change the motion of charged particles \cite{FrolovShoom:2010,AlievGaltsov:1989b}. The separability of the mentioned fundamental equations was generalized to a presence of such an additional weak electromagnetic field aligned along the primary Killing vector \cite{FrolovKrtous:2011,CarigliaEtal:2013b}.

The aim of this paper is to find the most general electromagnetic field admitting integrability of the charged particle motion and commutation of the charged scalar field operators in the \mbox{Kerr--NUT--(A)dS} spacetimes and related spaces. Then, for such a field, we demonstrate separations of the Hamilton--Jacobi and Klein--Gordon equations.

For this purpose we derive some general commutativity conditions of classical phase-space observables and scalar field operators. It is well known that conditions for vanishing Poisson brackets of classical observables are equivalent to the conditions of vanishing Schouten--Nijenhuis brackets \cite{Schouten:1940,Schouten:1954,Nijenhuis:1955}. However, commutativity of operators requires that additional anomalous conditions are satisfied. In the special case of the commutator of d'Alembertian with the second order operators, this was shown already in~\cite{Carter:1977}.

The paper is organized as follows. Section \ref{sc:comcond} deals with a general problem of the commutation of classical phase-space observables and their analogy for scalar field operators. We derive commutativity conditions for quadratic classical observables and second-order operators with an added electromagnetic field. In the case of operators, we obtain additional conditions. We discuss also the most interesting case when one element of the set of mutually commuting classical observables or operators represents Hamiltonian or d'Alembertian, respectively.

In Sec. \ref{sc:kerrnutads} we introduce \mbox{Kerr--NUT--(A)dS} spacetimes and spaces which share the same explicit and hidden symmetries. We find the most general form of the weak electromagnetic field preserving complete integrability of charged particle motion and the commutation of the operators. For such a field we solve the Hamilton--Jacobi and Klein--Gordon equations by separation of variables.

Section \ref{sc:concl} is devoted to conclusions. In the Appendix, we gather some useful identities which are used repeatedly in this paper.

\section{Commutativity conditions}\label{sc:comcond}

\subsection{Classical phase-space observables}\label{ssc:comcond-ob}

Particle motion in $D$-dimensional configuration space $M$ can be described in the language of Hamiltonian mechanics, where the phase space $S$ is the cotangent bundle $\bs{T}^*M$ equipped with the standard symplectic structure which induces the Poisson bracket. A phase-space point is described by a pair ${[x,\tb{p}]}$ with position ${x\in M}$ and momentum ${\tb{p}\in\bs{T}_{\!x}^*M}$. The manifold $M$ can represent either the generalization of a three-dimensional space in a nonrelativistic case or a four-dimensional spacetime in a relativistic case.\footnote{%
In the relativistic case the full spacetime $M$ plays a role of an extended configuration space and its cotangent bundle defines the extended phase space $S$. Hamiltonian formalism describes the evolution in particle inner time (time parameter along the trajectory). The expansion to the spacetime setting and the reparametrization freedom in inner time is compensated by the presence of the constraint (for the geodesic motion ${\tb{g}[^{ab}]\tb{p}[_a]\tb{p}[_b]+m^2=0}$) and by the appropriate gauge fixing (e.g, the choice of the proper time along the trajectory). Since the constraint is given by the observable which is conserved (actually, by the Hamiltonian itself), we do not have to concentrate on the constraint at the beginning. The constraint can be satisfied at the end just by choosing the correct value of the corresponding observable.}

The Poisson bracket of two observables ${A,\,B}$ can be written in terms of phase-space quantities as
\begin{equation}\label{eq:poisbr}
    \pois{A}{B} = \tb{\dd} A \con \tb{\Omega^{-1}{}} \con \tb{\dd} B\;,
\end{equation}
where $\tb{\Omega}$ is the standard symplectic structure \cite{Arnold:book,CarigliaEtal:2013a}. To write it down in terms of configuration space quantities we need to introduce an auxiliary structure. Namely, we assume that the manifold $M$ is endowed with a torsion-free covariant derivative\footnote{%
At this moment, the derivative $\tb{\nabla}$ is not necessary related to any metric. It could also be any coordinate derivative.}
$\tb{\nabla}$. Then, the Poisson bracket can be written \cite{CarigliaEtal:2013a} as\footnote{%
Here, $\frac{\pp }{\pp\smash{\tb{p}}}$ is a derivative with respect to the momentum at a fixed cotangent fiber $\bs{T}_{\!x}^*M$.  $\frac{\smash{\tb{\nabla}}}{\pp x}$ is the covariant differential with respect to the position, which ignores $\tb{p}$ dependence.}
\begin{equation}\label{eq:poisdef}
	\pois{A}{B}=\frac{\smash{\tb{\nablai}[_a]} A}{\pp x}\frac{\pp B}{\pp \smash{\tb{p}[_a]}}-\frac{\pp A}{\pp \smash{\tb{p}[_a]}}\frac{\smash{\tb{\nablai}[_a]} B}{\pp x}\;.
\end{equation}

If the observables are homogeneous powers of the momenta,
\begin{equation}\label{eq:homobs}
    A = \tb{A}[^{a_1\dots a_r}]\tb{p}[_{a_1}]\dots\tb{p}[_{a_r}]\;,\quad
    B = \tb{B}[^{a_1\dots a_s}]\tb{p}[_{a_1}]\dots\tb{p}[_{a_s}]
\end{equation}
(with the coefficients $\tb{A}$, $\tb{B}$ being rank-$r$ and rank-$s$ symmetric tensors), their Poisson bracket is also the homogeneous observable of type \eqref{eq:homobs} with the coefficient given by the Schouten--Nijenhuis bracket \cite{Schouten:1940,Schouten:1954,Nijenhuis:1955},
\begin{align}\label{eq:SNdef}
\begin{split}
	\sn{\tb{A}}{\tb{B}}[a_1\dots a_{r+s-1}]&\equiv r\tb{A}[^{b(a_1\dots a_{r-1}}]\tb{\nablai}[_b]\tb{B}[^{a_r\dots a_{r+s-1})}]\\
	&\feq-s\tb{B}[^{b(a_1\dots a_{s-1}}]\tb{\nablai}[_b]\tb{A}[^{a_s\dots a_{r+s-1})}]\;.
\end{split}
\end{align}
Let us stress, that although expressions \eqref{eq:poisdef} and \eqref{eq:SNdef} contain the derivative~$\tb{\nabla}$, the Poisson and Schouten--Nijenhuis  brackets themselves are independent of the choice of the derivative, cf.\ \cite{CarigliaEtal:2013a}.

A general classical phase-space observable $Q$, which is at most quadratic in the momentum $\tb{p}$, has the form
\begin{equation}\label{eq:quadobs}
	Q \equiv \tb{k}[^{ab}]\tb{p}[_a]\tb{p}[_b] + \tb{l}[^a]\tb{p}[_a]+v \;,
\end{equation}
where $\tb{k}$ is a rank-two symmetric tensor, $\tb{l}$ is a vector, and $v$ is a scalar. For further reference, we define homogeneous observables quadratic, linear, and constant in momenta,
\begin{equation}\label{eq:clasobs}
	K \equiv \tb{k}[^{ab}]\tb{p}[_a]\tb{p}[_b]\;,\quad
	L \equiv \tb{l}[^a]\tb{p}[_a]\;,\quad
	V \equiv v\;,
\end{equation}
i.e., ${Q=K+L+V}$.

Consider a set of the classical observables $K_i$, $L_j$ of the form \eqref{eq:clasobs}. The Poisson brackets of these observables may be written as\footnote{%
Here, $\tb{p^3}$ denotes the tensorial power, $\tb{\pow{p}{3}}[_{abc}] = \tb{p}[_a]\,\tb{p}[_b]\,\tb{p}[_c]$. Similarly, in the next section, $\tb{\pow{A}{2}}[_{ab}] = \tb{A}[_a]\tb{A}[_b]$.}
\begin{equation} \label{eq:VLKpois}
\begin{aligned}
	\pois{K_i}{K_j} &=-\sn{\tb[^{\ind{i}}]{k}}{\tb[^{\ind{j}}]{k}}[abc]\;\tb{\pow{p}{3}}[_{abc}]\;,\\
	\pois{L_i}{K_j} &=-\sn{\tb[^{\ind{i}}]{l}}{\tb[^{\ind{j}}]{k}}[ab]\;\tb{\pow{p}{2}}[_{ab}]\;,\\
	\pois{L_i}{L_j} &=-\sn{\tb[^{\ind{i}}]{l}}{\tb[^{\ind{j}}]{l}}[a]\; \tb{p}[_a]\;
\end{aligned}
\end{equation}
and the Poisson brackets with the observable $V$ independent of momenta as
\begin{equation}\label{eq:VLKpoisV}
\begin{aligned}
	\pois{V}{K_i} &=-\sn{v}{\tb[^{\ind{i}}]{k}}[a]\;\tb{p}[_a]\;,\\
	\pois{V}{L_i} &=-\sn{v}{\tb[^{\ind{i}}]{l}}\;.
\end{aligned}
\end{equation}

We thus see from \eqref{eq:VLKpois} that $K_i$, $L_j$ mutually Poisson-commute (are in involution) if and only if the corresponding Schouten--Nijenhuis brackets vanish
\begin{align}	
	\sn{\tb[^{\ind{i}}]{k}}{\tb[^{\ind{j}}]{k}} &=0\;, \label{eq:obmutualcom1}\\
	\sn{\tb[^{\ind{i}}]{l}}{\tb[^{\ind{j}}]{k}} &=0\;, \label{eq:obmutualcom2}\\
	\sn{\tb[^{\ind{i}}]{l}}{\tb[^{\ind{j}}]{l}} &=0\;. \label{eq:obmutualcom3}
\end{align}

\subsection{Scalar field operators}\label{ssc:comcond-op}

Now we would like to study an operator analogy of the classical observables discussed above. Applying the standard heuristic rule ${\tb{p}\rightarrow-\imag\tb{\nabla}}$ and employing symmetric operator ordering, the quadratic observable \eqref{eq:quadobs} yields a general second-order Hermitian operator $\op{Q}$ of the form\footnote{%
In operator equations we use the convention that the round brackets around a derivative end the action of the derivative to the right, however, the square brackets do not. It means ${\big[\tb{\nablai}[_a]\tb{l}[^a]\big] = \tb{l}[^a]\tb{\nablai}[_a] + \big(\tb{\nablai}[_a]\tb{l}[^a]\big)}$. Applying the operator on a scalar ${\phi}$ we get ${\big[\tb{\nablai}[_a]\tb{l}[^a]\big]\phi = \big(\tb{\nablai}[_a](\tb{l}[^a]\phi)\big) = \tb{l}[^a]\big(\tb{\nablai}[_a]\phi\big) + \phi \big(\tb{\nablai}[_a]\tb{l}[^a]\big)}$.}
\begin{equation}\label{eq:ghermop}
\begin{aligned}
  \op{Q}
   &\equiv-\tb{\nablai}[_a]\tb{k}[^{ab}]\tb{\nablai}[_b]-\frac{\imag}{2}\big[\tb{l}[^a]\tb{\nablai}[_a]+\tb{\nablai}[_a]\tb{l}[^a]\big]+v\\
   &=-\tb{k}[^{ab}]\tb{\nablan{2}}[_{ab}]-\big(\tb{\nablai}[_a]\tb{k}[^{ab}]{+}\,\imag\tb{l}[^b]\big)\tb{\nablai}[_b]
     -\frac{\imag}{2}\big(\tb{\nablai}[_a]\tb{l}[^a]\big) + v\;.
\end{aligned}
\end{equation}
Here, we introduce the symmetric higher-order covariant derivative
\begin{equation}\label{eq:symcd}
    \tb{\nablan{r}}[_{a_1\dots a_r}] = \tb{\nablai}[_{(a_1}]\dots \tb{\nablai}[_{a_r)}]\;.
\end{equation}

When applying the rule ${\tb{p}\rightarrow-\imag\tb{\nabla}}$, we have to choose the order of $x$ dependent tensorial coefficients and derivatives. As we said, we used the symmetric ordering as can be seen in the first expression of \eqref{eq:ghermop}. The second expression is an equivalent ``normal'' form when the derivatives are placed on the right. Other operator orderings would differ from the symmetric one by terms of lower derivative orders. Therefore, we can restrict ourselves to the operator of the form above, remembering that the operator coefficients of lower order can be, in principle, different from those in the classical version of the observable.

Even if we fix the operator ordering, the operators do not depend just on the tensorial coefficients as the classical observables, but also on the choice of the covariant derivative $\tb{\nabla}$. We do not assume at this moment that it has to be the Levi-Civita derivative, although, it will be a typical choice. But we assume that it is a torsion-free derivative flat on densities, i.e., satisfying ${\tb{R}[_{ab}^{n}_{n}]=0}$, which is equivalent to
\begin{align} \label{eq:riccisym}
\tb{\Ric}[_{ab}]=\tb{\Ric}[_{(ab)}]\;.
\end{align}
Since, two such covariant derivatives differ only by a tensorial term, the corresponding operators would differ only in lower derivative terms. The freedom in the choice of the derivative is thus similar to the freedom of the operator ordering. A particular choice of the covariant derivative thus does not change a whole class of operators of a given order, it only changes how we parameterize operators in the class. Moreover, the highest order coefficient (the principal symbol) is independent of the choice of the derivative.

Similarly to the classical case \eqref{eq:clasobs}, we introduce homogeneous Hermitian operators of the second order, the first order, and the zeroth order, respectively,
\begin{align}\label{eq:quantop}
\begin{split}
  \op{K}
    &\equiv -\tb{\nablai}[_a]\tb{k}[^{ab}]\tb{\nablai}[_b]
    =-\tb{k}[^{ab}]\tb{\nablan{2}}[_{ab}]-\big(\tb{\nablai}[_a]\tb{k}[^{ab}]\big)\tb{\nablai}[_b]\;,\\
  \op{L}
    &\equiv-\frac{\imag}{2}\big[\tb{l}[^a]\tb{\nablai}[_a]+\tb{\nablai}[_a]\tb{l}[^a]\big]
    =-\imag\,\tb{l}[^a]\tb{\nablai}[_a] -\frac{\imag}{2}\big(\tb{\nablai}[_a]\tb{l}[^a]\big)\;,\\
  \op{V} &\equiv  v\;,
\end{split}
\end{align}
i.e., ${{\op{Q}}={\op{K}}+{\op{L}}+{\op{V}}}$.

Let us now consider a set of operators ${\op{K}}_i$, ${\op{L}}_j$ of the form \eqref{eq:quantop}. A rather involved calculation [using identities \eqref{eq:symkov} from the Appendix and \eqref{eq:riccisym}] shows that the commutators of these operators may be written as
\begin{align}\label{eq:VLKcom}
\begin{split}
  \com{{\op{K}}_i}{{\op{K}}_j} &=\sn{\tb[^{\ind{i}}]{k}}{\tb[^{\ind{j}}]{k}}[abc]\,\tb{\nablan{3}}[_{abc}]
    +\frac{3}{2}\big(\tb{\nablai}[_c]\sn{\tb[^{\ind{i}}]{k}}{\tb[^{\ind{j}}]{k}}[abc]\big)\tb{{\nablan{2}}}[_{ab}]
    \\	
    &\feq+\frac{1}{2}\Big(\tb{\nablai}[_a]\big(\tb{\nablai}[_c]\sn{\tb[^{\ind{i}}]{k}}{\tb[^{\ind{j}}]{k}}[abc]+\tb[^{\ind{ij}}]{m}[^{ab}]\big)\Big)\tb{{\nablai}}[_b]\;,
    \\	
  \com{{\op{L}}_i}{{\op{K}}_j} &=\imag\sn{\tb[^{\ind{i}}]{l}}{\tb[^{\ind{j}}]{k}}[ab] \tb{\nablan{2}}[_{ab}]
    +\imag\big(\tb{\nablai}[_a]\sn{\tb[^{\ind{i}}]{l}}{\tb[^{\ind{j}}]{k}}[ab]\big)\tb{\nablai}[_b]
    \\
	&\feq -\frac{\imag}{2}\Big(\tb{\nablai}[_a]\big(\tb[^{\ind{j}}]{k}[^{ab}]\tb{\nablai}[_b](\tb{\nablai}[_c]\tb[^{\ind{i}}]{l}[^c])\big)\Big)\;,
    \\
  \com{{\op{L}}_i}{{\op{L}}_j} &=-\sn{\tb[^{\ind{i}}]{l}}{\tb[^{\ind{j}}]{l}}[a]\tb{{\nablai}}[_a]
    - \frac{1}{2}\big(\tb{\nablai}[_a]\sn{\tb[^{\ind{i}}]{l}}{\tb[^{\ind{j}}]{l}}[a]\big)\;,
\end{split}
\end{align}
and commutators with zeroth order operator $\op{V}$ as
\begin{equation}\label{eq:VLKcomV}
\begin{split}
  \com{\op{V}}{{\op{K}}_i} &= -\sn{v}{\tb[^{\ind{i}}]{k}}[a]\tb{{\nablai}}[_a] -\frac{1}{2}\big(\tb{\nablai}[_a] \sn{v}{\tb[^{\ind{i}}]{k}}[a]\big)\;,
\\
  \com{\op{V}}{{\op{L}}_i} &=-\imag\sn{v}{\tb[^{\ind{i}}]{l}}\;.
\end{split}
\end{equation}
Here, $\tb[^{\ind{ij}}]{m}$ is an antisymmetric tensor defined by
\begin{align}
\begin{split}
  \tb[^{\ind{ij}}]{m}[^{ab}]
    &\equiv \frac{2}{3}\Big(\tb[^{\ind{i}}]{k}[^{c[a}]\tb{\nablai}[_d]\tb{\nablai}[_c]\tb[^{\ind{j}}]{k}[^{b]d}]-\tb[^{\ind{j}}]{k}[^{c[a}]\tb{\nablai}[_d]\tb{\nablai}[_c]\tb[^{\ind{i}}]{k}[^{b]d}]\Big)
    \\
	 &\feq-\frac{2}{3}\big(\tb{\nablai}[_d]\tb[^{\ind{i}}]{k}[^{c[a}]\big)\big(\tb{\nablai}[_c]\tb[^{\ind{j}}]{k}[^{b]d}]\big)-2\tb[^{\ind{i}}]{k}[^{c[a}]\tb{\Ric}[_{cd}]\tb[^{\ind{j}}]{k}[^{b]d}]\;.
\end{split}\raisetag{9ex}
\end{align}

Checking the coefficients in front of all symmetric derivatives in \eqref{eq:VLKcom} we find that the operators ${\op{K}}_i$, ${\op{L}}_j$ mutually commute if and only if \eqref{eq:obmutualcom1}--\eqref{eq:obmutualcom3} hold and, in contrast to the commutation of the classical observables, additional conditions
\begin{gather}
  \tb{\nablai}[_a]\big(\tb[^{\ind{i}}]{k}[^{ab}]\tb{\nablai}[_b](\tb{\nablai}[_c]\tb[^{\ind{j}}]{l}[^c])\big) =0\;,
    \label{eq:opmutualcom4}\\
  \tb{\nablai}[_a]\tb[^{\ind{ij}}]{m}[^{ab}] =0\;
    \label{eq:opmutualcom5}
\end{gather}
must be also met.

We call these conditions \emph{anomalous conditions}. Depending on the choice of the covariant derivative $\tb{\nabla}$, some of these conditions can be consequences of the classical conditions \eqref{eq:obmutualcom1}--\eqref{eq:obmutualcom3}; but, in general, they constitute new nontrivial requirements on the operator coefficients, $\tb[^{\ind{i}}]{k}$ and $\tb[^{\ind{j}}]{l}$.

\subsection{Charged classical observables}\label{ssc:comcond-chob}

Besides the geodesic motion, it is interesting to study also particle motion in an electromagnetic field. To obtain a system with conserved quantities in involution in such a case, the symmetries of the background geometry have to be accompanied by an analogous symmetry of the electromagnetic field. In this section we find such general compatibility conditions.

We define ``charged'' classical observables as follows:
\begin{align}\label{eq:clasobch}
\begin{split}
  \ti{q}K &\equiv \big(\tb{p}[_a]-q\tb{A}[_a]\big) \tb{k}[^{ab}] \big(\tb{p}[_b]-q\tb{A}[_b]\big)
    \\
  &=K-2q\tb{A}[_a]\tb{k}[^{ab}]\tb{p}[_b] +q^2\tb{\pow{A}{2}}[_{ab}]\,\tb{k}[^{ab}]\;,
    \\
  \ti{q}L &\equiv \tb{l}[^a]\tb{p}[_a] = L\;,
\end{split}
\end{align}
where $\tb{A}$ is the vector potential describing the electromagnetic field and $q$ is a charge of the particle. $K$ and $L$ represent the uncharged classical observables \eqref{eq:clasobs} defined above.

Consider a set\footnote{The indices ${i,j,\dots}$ labels the observables and their tensorial coefficients $\tb{k}$ and $\tb{l}$. They are not tensor indices and therefore they do not enter tensorial operations in various expressions below.} of classical observables $\ti{q}K_i$, $\ti{q}L_j$, now of the form \eqref{eq:clasobch}. With the help of relations \eqref{eq:VLKpois}, \eqref{eq:VLKpoisV}, and \eqref{eq:SNidentities}, the Poisson brackets of these classical observables may be written in the form
\begin{equation}\label{eq:VLKpoisch}
\begin{split}
  \pois{\ti{q}K_i}{\ti{q}K_j}
    &=\pois{K_i}{K_j}
    \\
  \shiftandleft{2.5em}+q\Big(3\tb{A}[_c]\,\sn{\tb[^{\ind{i}}]{k}}{\tb[^{\ind{j}}]{k}}[abc]
    +4\tb[^{\ind{i}}]{k}[^{c(a}]\tb{F}[_{cd}]\tb[^{\ind{j}}]{k}[^{b)d}]\Big)
    \,\tb{\pow{p}{2}}[_{ab}]
    \\
  \shiftandleft{2.5em}-q^2\Big(3\tb{\pow{A}{2}}[_{bc}]\,\sn{\tb[^{\ind{i}}]{k}}{\tb[^{\ind{j}}]{k}}[abc]
    +8\tb{A}[_b]\tb[^{\ind{i}}]{k}[^{c(a}]\tb{F}[_{cd}]\tb[^{\ind{j}}]{k}[^{b)d}]\Big)\,\tb{p}[_a]
  	\\
  \shiftandleft{2.5em}+q^3\Big(\tb{\pow{A}{3}}[_{abc}]\,\sn{\tb[^{\ind{i}}]{k}}{\tb[^{\ind{j}}]{k}}[abc]
    +4\tb{\pow{A}{2}}[_{ab}]\tb[^{\ind{i}}]{k}[^{c(a}]\tb{F}[_{cd}]\tb[^{\ind{j}}]{k}[^{b)d}]\Big)\;,
    \\
  \pois{\ti{q}L_i}{\ti{q}K_j}
    &=\pois{L_i}{K_j}
    \\
  \shiftandleft{2.5em}+ 2q\Big(\tb{A}[_a]\,\sn{\tb[^{\ind{i}}]{l}}{\tb[^{\ind{j}}]{k}}[ab]
    +\big(\lie{\tb[^{\ind{i}}]{l}}\,\tb{A}[_a]\big)\tb[^{\ind{j}}]{k}[^{ab}]\Big)\,\tb{p}[_b]
    \\
  \shiftandleft{2.5em}-q^2\Big(\tb{\pow{A}{2}}[_{ab}]\,\sn{\tb[^{\ind{i}}]{l}}{\tb[^{\ind{j}}]{k}}[ab]
    +2\big(\lie{\tb[^{\ind{i}}]{l}}\,\tb{A}[_a]\big)\tb[^{\ind{j}}]{k}[^{ab}]\tb{A}[_b]\Big)\;,
    \\
  \pois{\ti{q}L_i}{\ti{q}L_j} &=\pois{L_i}{L_j}\;.
\end{split}\raisetag{17ex}
\end{equation}\nopagebreak
\pagebreak[4]
Here, ${\tb{F}\equiv\tb{\dd}\tb{A}}$ denotes the Maxwell tensor.

The requirement that the observables $\ti{q}K_i$, $\ti{q}L_j$ mutually Poisson commute gives us conditions both on the tensor coefficients  $\tb[^{\ind{i}}]{k}$, $\tb[^{\ind{j}}]{l}$, as well as on the vector potential $\tb{A}$.
Reading out different powers in the charge, it follows from \eqref{eq:VLKpois} that ``uncharged'' conditions  \eqref{eq:obmutualcom1}--\eqref{eq:obmutualcom3} must still hold, and, additionally, the field $\tb{A}$ has to satisfy
\begin{gather}
  \lie{\tb[^{\ind{i}}]{l}}\,{\tb{A}} =0\;,
    \label{eq:Aobcond1}\\
  \tb[^{\ind{i}}]{k}[^{c(a}]\tb{F}[_{cd}] \tb[^{\ind{j}}]{k}[^{b)d}] =0\;.
    \label{eq:Aobcond2}
\end{gather}

The first condition has the clear meaning: the vector potential cannot change along the explicit symmetries given by the generators $\tb[^{\ind{i}}]{l}$. The second condition expresses the compatibility with the hidden symmetries. It has an algebraic form in terms of the Maxwell tensor~$\tb{F}$.


\subsection{Charged field operators}\label{ssc:comcond-chop}

Similarly to the uncharged case, we define operator analogues of the classical charged observables \eqref{eq:clasobch}
\begin{align}\label{eq:opch}
\begin{split}
  \ti{q\,}{\op{K}} &\equiv
    -\big[\tb{{\nablai}}[_a]-\imag q \tb{A}[_a]\big] \tb{k}[^{ab}]
    \big[\tb{{\nablai}}[_b]-\imag q \tb{A}[_b]\big]
     \\
  &={\op{K}}+2\imag q\tb{A}[_a]\tb{k}[^{ab}]\tb{{\nablai}}[_b]
    +\imag q \big(\tb{\nablai}[_a](\tb{k}[^{ab}]\tb{A}[_b])\big)
    +q^2\tb{\pow{A}{2}}[_{ab}]\,\tb{k}[^{ab}]\;,
    \\
  \ti{q\,}{\op{L}} &\equiv -\frac{\imag}{2}\big[\tb{l}[^a]\tb{\nablai}[_a]+\tb{\nablai}[_a]\tb{l}[^a]\big]={\op{L}}\;,
\end{split}\raisetag{4ex}
\end{align}
where ${\op{K}}$ and ${\op{L}}$ are uncharged operators \eqref{eq:quantop}.

Likewise, consider a set of the field operators $\ti{q\,}{\op{K}}_i$, $\ti{q\,}{\op{K}}_j$ of the form \eqref{eq:opch}. Employing relations \eqref{eq:VLKcom}, \eqref{eq:VLKcomV} and \eqref{eq:SNidentities}, commutators of these charged operators may be written in the form
\begin{align}\label{eq:VLKcomch}
\begin{split}
 \com{\ti{q\,}{\op{K}}_i}{\ti{q\,}{\op{K}}_j}
    &= \com{{\op{K}}_i}{{\op{K}}_j}
    \\
    \shiftandleft{2.5em} -\imag q\Big(3\tb{A}[_c]\,\sn{\tb[^{\ind{i}}]{k}}{\tb[^{\ind{j}}]{k}}[abc]
    +4\tb[^{\ind{i}}]{k}[^{c(a}]\tb{F}[_{cd}] \tb[^{\ind{j}}]{k}[^{b)d}]\Big)\,
    \tb{\nablan{2}}[_{ab}]
    \\
  \shiftandleft{2.5em} -\bigg(
    \imag q\tb{\nablai}[_b]\Big(3\tb{A}[_c]\,\sn{\tb[^{\ind{i}}]{k}}{\tb[^{\ind{j}}]{k}}[abc]
    +4\tb[^{\ind{i}}]{k}[^{c(a}]\tb{F}[_{cd}] \tb[^{\ind{j}}]{k}[^{b)d}]\Big)
    \\
  \shiftandleft{2.5em} +q^2\Big(3\tb{\pow{A}{2}}[_{bc}]\,\sn{\tb[^{\ind{i}}]{k}}{\tb[^{\ind{j}}]{k}}[abc]
    +8\tb{A}[_b]\tb[^{\ind{i}}]{k}[^{c(a}]\tb{F}[_{cd}] \tb[^{\ind{j}}]{k}[^{b)d}]\Big)
    \bigg)\tb{{\nablai}}[_a]
    \\
  \shiftandleft{2.5em} +\bigg(
    \imag q^3 \Big(\tb{\pow{A}{3}}[_{abc}]\,\sn{\tb[^{\ind{i}}]{k}}{\tb[^{\ind{j}}]{k}}[abc]
    + 4\tb{\pow{A}{2}}[_{ab}]\tb[^{\ind{i}}]{k}[^{c(a}]\tb{F}[_{cd}] \tb[^{\ind{j}}]{k}[^{b)d}]\Big)
    \\
  \shiftandleft{2.5em} -\frac{1}{2}q^2\tb{\nablai}[_a]\Big(3\tb{\pow{A}{2}}[_{bc}]\,\sn{\tb[^{\ind{i}}]{k}}{\tb[^{\ind{j}}]{k}}[abc]
    +8\tb{A}[_b]\tb[^{\ind{i}}]{k}[^{c(a}]\tb{F}[_{cd}]\tb[^{\ind{j}}]{k}[^{b)d}]\Big)
    \\
  \shiftandleft{2.5em} +\imag q\tb{\nablai}[_a]\Big(
    \tb[^{\ind{j}}]{k}[^{ab}]\tb{\nablai}[_b]\big(\tb{\nablai}[_c](\tb[^{\ind{i}}]{k}[^{cd}]\tb{A}[_d])\big)
    -\switch{i}{j}\Big)\bigg)\;,
    \\
  \com{\ti{q\,}{\op{L}}_i}{\ti{q\,}{\op{K}}_j} &= \com{{\op{L}}_i}{{\op{K}}_j}
  	\\
  \shiftandleft{2.5em} +2q\big(\tb{A}[_a]\sn{\tb[^{\ind{i}}]{l}}{\tb[^{\ind{j}}]{k}}[ab]
    +\big(\lie{\tb[^{\ind{i}}]{l}}\,{\tb{A}[_a]}\big)\tb[^{\ind{j}}]{k}[^{ab}]\big)\tb{{\nablai}}[_b]
    \\
  \shiftandleft{2.5em} +q\Big(\tb{\nablai}[_b]\big(\tb{A}[_a]\sn{\tb[^{\ind{i}}]{l}}{\tb[^{\ind{j}}]{k}}[ab]
    +\big(\lie{\tb[^{\ind{i}}]{l}}\,{\tb{A}[_a]}\big)\tb[^{\ind{j}}]{k}[^{ab}]\big)\Big)
   \\
  \shiftandleft{2.5em} -\imag q^2\big(\tb{\pow{A}{2}}[_{ab}]\,\sn{\tb[^{\ind{i}}]{l}}{\tb[^{\ind{j}}]{k}}[ab]
    +2\big(\lie{\tb[^{\ind{i}}]{l}}\,{\tb{A}[_a]}\big)\tb[^{\ind{j}}]{k}[^{ab}]\tb{A}[_b]\big)\;,
    \\
  \com{\ti{q\,}{\op{L}}_i}{\ti{q\,}{\op{L}}_j} &= \com{{\op{L}}_i}{{\op{L}}_j}\;.
\end{split}\raisetag{26ex}
\end{align}

Requiring the mutual commutativity for the operators $\ti{q\,}{\op{K}}_i$, $\ti{q\,}{\op{L}}_j$ we find that, in addition to the ``uncharged'' conditions \eqref{eq:obmutualcom1}--\eqref{eq:obmutualcom3}, \eqref{eq:opmutualcom4}, and \eqref{eq:opmutualcom5}, we need the field $\tb{A}$ to satisfy the ``classical'' conditions \eqref{eq:Aobcond1}, \eqref{eq:Aobcond2}, and the new anomalous condition
\begin{equation}\label{eq:Aopcond3}
  \tb{\nablai}[_a]\Big(\tb[^{\ind{i}}]{k}[^{ab}]\tb{\nablai}[_b]
    \big(\tb{\nablai}[_c](\tb[^{\ind{j}}]{k}[^{cd}]\tb{A}[_d])\big)-\switch{i}{j}\Big) =0\;.
\end{equation}
Notice that this condition depends on the choice of the covariant derivative.

\subsection{Conserved quantities}\label{ssc:comcond-cq}

Let us concentrate on a physical situation when the configuration space is endowed with a particular geometry given by the metric $\tb{g}$. The motion of a relativistic as well as a nonrelativistic particle in this spacetime can be described by a Hamiltonian which is quadratic in momentum\footnote{%
We ignore a possible multiplicative constant which affects only a scale of the time parameter.}
\begin{equation}
	H\equiv\tb{g}[^{ab}]\tb{\pow{p}{2}}[_{ab}]\;.
\end{equation}
We are interested in the case when there exists a set of independent mutually Poisson-commuting classical observables $K_i$, $L_j$, of the form \eqref{eq:clasobs} and $H$ is one of the observables $K_i$. It implies that all observables ${K_i,\,L_j}$ are conserved quantities.

The conditions \eqref{eq:obmutualcom1} and \eqref{eq:obmutualcom2} of the Poisson commutation with the Hamiltonian then imply that  $\tb[^{\ind{i}}]{k}$ and $\tb[^{\ind{j}}]{l}$ must be the Killing tensors and vectors of the metric~$\tb{g}$, respectively. Indeed, choosing the covariant derivative ${\tb{\nabla}}$ associated with the metric ${\tb{g}}$, these conditions become
\begin{equation}\label{eq:killeq}
\begin{aligned}
	\tb{\nabla}[^{(a}]\tb[^{\ind{i}}]{k}[^{bc)}]=0\;, \quad \tb{\nabla}[^{(a}]\tb[^{\ind{i}}]{l}[^{b)}]=0\;.
\end{aligned}
\end{equation}
Here, the indices are raised using the metric $\tb{g}$.

Similarly, we define the d'Alembertian
\begin{equation}
	\Box\equiv\tb{g}[^{ab}]\tb{\nablan{2}}[_{ab}]
\end{equation}
and we assume a set of independent mutually commuting operators  ${\op{K}}_i$, ${\op{L}}_j$ of the form \eqref{eq:quantop}, where the d'Alembertian $\Box$ is one of the operators ${\op{K}}_i$. Then, the tensorial coefficients $\tb[^{\ind{i}}]{k}$ and $\tb[^{\ind{i}}]{l}$ must be Killing tensors and vectors of the metric $\tb{g}$, respectively, and additionally, the anomalous conditions \eqref{eq:opmutualcom4} and \eqref{eq:opmutualcom5}, must be satisfied. Those anomalous conditions, which arise from the commutation with $\Box$, thanks to \eqref{eq:riccisym} and \eqref{eq:killeq}, simplify to\footnote{This anomalous condition has been already derived in \cite{Carter:1977}.}
\begin{equation}\label{eq:anomric}
	\tb{\nablai}[_a]\Big(\tb{\Ric}[^a_c]\tb[^{\ind{i}}]{k}[^c_b]
    -\tb[^{\ind{i}}]{k}[^a_c]\tb{\Ric}[^c_b]\Big)=0\;.
\end{equation}
The indices are raised again by the metric $\tb{g}$.

Condition \eqref{eq:anomric} (for a given ${i}$) is automatically met in several situations. For instance, it is satisfied if $\tb{g}$ solves vacuum Einstein equations (admitting the cosmological constant), or if the Ricci tensor and the Killing tensor can be simultaneously diagonalized, or if the Killing tensor can be written as ${\tb[^{\ind{i}}]{k}[^a_b]=\tb{f}[^a_c]\tb{f}[^c_b]}$ for a Killing--Yano \mbox{2-form}~$\tb{f}$.

The last statement, mentioned in \cite{Cariglia:2004}, can be generalized to the case when the Killing tensor is a square of the Killing--Yano ${p}$-form generated from the principal closed conformal Killing--Yano tensor. The existence of such a structure leads us, thanks to the uniqueness result \cite{KrtousFrolovKubiznak:2008}, to \mbox{Kerr--NUT--(A)dS} spacetimes.

\section{Kerr--NUT--(A)dS and related spaces}\label{sc:kerrnutads}

\subsection{Kerr--NUT--(A)dS metric}\label{scc:kerrnutads-kerrnutads}

\mbox{Kerr--NUT--(A)dS} spacetimes \cite{ChenLuPope:2006} in an even dimension\footnote{%
We restrict ourselves to even dimensions mainly because of the simplicity of equations. All results can be generalized to odd dimensions, when some of the expressions below include additional odd terms.}
 ${D=2n}$ are given by the metric\footnote{%
In the following, we do not assume an implicit sum over greek indices ${\mu,\nu,\dots}$ and latin indices ${i,j,\dots}$. Unless otherwise stated, indices have ranges ${\mu,\nu = 1,\dots,n}$ and ${i,j=0,\dots,n-1}$, respectively. We use shortened notations ${\sum_\mu \equiv \sum_{\mu=1}^n}$, ${\sum_i \equiv \sum_{i=0}^{n-1}}$ and ${\prod_\mu \equiv \prod_{\mu=1}^n}$, ${\prod_i \equiv \prod_{i=0}^{n-1}}$.}
\begin{align}
  \tb{g}=\sum_{\mu}\bigg[\frac{U_\mu}{X_\mu}\big(\tb{\dd}x_\mu\big)^{\!\bs{2}}
  +\frac{X_\mu}{U_\mu}\Big(\sum_{j}A_{\mu}^{(j)}\tb{\dd}\psi_j\Big)^{\!\bs{2}}\bigg]\;. \label{eq:kerrnutads}
\end{align}
Coordinates $x_\mu$, ${\mu=1,\dots,n}$, correspond to the radial and azimuthal directions and Killing coordinates $\psi_j$, ${j=0,\dots,n-1}$, denote the temporal and rotational directions. The metric is written in a Euclidean form in which the radial coordinate and some other quantities are multiplied by the imaginary unit $\imag$ in order to achieve a more symmetric form. However, for a suitable choice of ranges of coordinates, the metric is real and has a Lorentzian signature ${(-++\cdots)}$ \cite{ChenLuPope:2006,Krtous:inprep2015}.

Functions $U_\mu$ and $A_{\mu}^{(i)}$ are defined by
\begin{align}\label{eq:AUfunctions}
\begin{split}
	U_\mu &\equiv \prod_{\mathclap{\substack{\nu\\ \nu\neq\mu}}} \big(x_\nu^2-x_\mu^2\big)\;, \\
	A_{\mu}^{(i)} &\equiv \sum_{\mathclap{\substack{\nu_1,\dots,\nu_i \\ \nu_1<\dots<\nu_i \\ \nu_k\neq\mu}}} x_{\nu_1}^2 \dots x_{\nu_i}^2\;.
\end{split}
\end{align}
In order to satisfy vacuum Einstein equations, the metric function ${X_\mu}$ must have the form
\begin{align}
	X_\mu=b_\mu x_\mu+\sum_{k=0}^{n}c_k x_\mu^{2k}\;,
\end{align}
where the constants $c_k$ and $b_\mu$ relate to the cosmological constant, angular momenta, mass, and NUT charges of a black hole. However, in the following, we can consider the so-called off-shell metric, when we just assume that ${X_\mu=X_\mu(x_\mu)}$. It means that each $X_\mu$ is an arbitrary function of a single variable $x_\mu$. Our conclusions do not depend on particular forms of these functions.

We introduce an un-normalized orthogonal vector frame ${\epsv{\mu},\,\epsvh{\mu}}$, related normalized orthonormal frame ${\ev{\mu},\,\evh{\mu}}$, and the dual covector frames ${\epsf{\mu},\,\epsfh{\mu}}$ and ${\ef{\mu},\,\efh{\mu}}$,
\begin{align}\label{eq:e&epsbase}
\begin{split}
	\epsv{\mu} &=\frac{1}{\sqrt{Q_\mu}}\ev{\mu}=\tvec{x_\mu}\;,\\
	\epsvh{\mu} &=\sqrt{Q_\mu}\evh{\mu}
       =\sum_{k}\frac{\big(\!-\!x_\mu^2\big)^{\!n-1-k}}{U_\mu}\tvec{\psi_k}\;,
\\
	\epsf{\mu} &=\sqrt{Q_\mu}\ef{\mu}=\tb{\dd}x_\mu\;,\\
	\epsfh{\mu}&=\frac{1}{\sqrt{Q_\mu}}\efh{\mu}=\sum_{k}A_{\mu}^{(k)}\tb{\dd}\psi_k\;.
\end{split}
\end{align}
where
\begin{equation}\label{eq:Qdef}
    Q_\mu \equiv \frac{X_\mu}{U_\mu}\;.
\end{equation}

It was shown in \cite{KrtousEtal:2007a} that the geometry \eqref{eq:kerrnutads} possess hidden and explicit symmetries encoded by rank-two Killing tensors $\tb[^{\ind{i}}]{k}$ and Killing vectors $\tb[^{\ind{i}}]{l}$, respectively,
\begin{align}
  \tb[^{\ind{i}}]{k}
    &\equiv\sum_{\mu}A_{\mu}^{(i)}\!\bigg[Q_\mu\bigg(\!\tvec{x_\mu}\!\bigg)^{\!\!\bs{2}}
    {+}\frac{1}{Q_\mu}\!\bigg(\!\!\sum_{k}\frac{\big({-}x_\mu^2\big)^{\!n{-}1{-}k}}{U_\mu}\tvec{\psi_k}\!\bigg)^{\!\!\bs{2}}\bigg]\nonumber\\
  &=\sum_{\mu}A_{\mu}^{(i)}\big(\ev{\mu}\ev{\mu}+\evh{\mu}\evh{\mu}\big)\;,\label{eq:klinbase}\\
	\tb[^{\ind{i}}]{l} &\equiv\tvec{\psi_i}=\sum_\mu A_{\mu}^{(i)}\sqrt{Q_\mu}\evh{\mu}\;.\nonumber
\end{align}
The corresponding classical observables
\begin{align}
\begin{split}
	K_i =\tb[^{\ind{i}}]{k}[^{ab}]\tb{\pow{p}{2}}[_{ab}]\;,\quad	L_i =\tb[^{\ind{i}}]{l}[^a]\tb{p}[_a]
\end{split}
\end{align}
mutually Poisson commute \cite{KrtousEtal:2007b} and the quantities $\tb[^{\ind{i}}]{k}$ and $\tb[^{\ind{i}}]{l}$ satisfy \eqref{eq:obmutualcom1}--\eqref{eq:obmutualcom3}. The Hamiltonian for geodesic motion is given by ${H=K_0}$, i.e., ${\tb[^{\ind{0}}]{k}}$ is inverse of the metric \eqref{eq:kerrnutads}. Moreover, all Killing tensors ${ \tb[^{\ind{i}}]{k}}$ can be written as a contracted square of Killing--Yano forms \cite{KrtousEtal:2007b}.

As can be observed from \eqref{eq:klinbase}, the frames \eqref{eq:e&epsbase} are formed by common eigenvectors of all Killing tensors ${\tb[^{\ind{i}}]{k}}$. It turns out that the Ricci tensor can be diagonalized in the same frame \cite{HamamotoEtal:2007}, which implies that the anomalous conditions \eqref{eq:anomric} are satisfied. Actually, the scalar field operators ${\op{K}_i,\,\op{L}_j}$ given by \eqref{eq:opch} employing the standard metric derivative ${\tb{\nabla}}$ mutually commute. Direct proof of this property, based on the coordinate expressions for the operators, has been presented in \cite{SergyeyevKrtous:2008}. We can thus conclude that all anomalous conditions \eqref{eq:opmutualcom4} and \eqref{eq:opmutualcom5} are satisfied.


\subsection{Geometries related to Kerr--NUT--(A)dS spaces}\label{scc:kerrnutads-rel}

We can define a set of new metrics inverting each tensor~${\tb[^{\ind{i}}]{k}}$,
\begin{align}\label{eq:igmetrics}
  \tb[^{\ind{i}}]{g}\equiv\tb[^{\ind{i}}]{k^{-1}{}}
  =\sum_{\mu}\frac{1}{A_{\mu}^{(i)}}\big(\ef{\mu}\ef{\mu}+\efh{\mu}\efh{\mu}\big)\;.
\end{align}
With each of these metrics we associate the corresponding metric covariant derivative $\tb[^{\ind{i}}]{\nabla}$, i.e., a torsion-free covariant derivative satisfying ${\tb[^{\ind{i}}]{\nabla}\tb[^{\ind{i}}]{g}=0}$. The metric $\tb[^{\ind{0}}]{g}$ is the original metric ${\tb{g}}$ of the off-shell \mbox{Kerr--NUT--(A)dS} spacetimes. Unfortunately, contrary to the Kerr--NUT--AdS, spacetime metrics \eqref{eq:igmetrics} with ${i>0}$ do not solve vacuum Einstein equations for any functions $X_\mu$. It also seems that these metrics do not admit a principal closed conformal Killing--Yano tensor.\footnote{%
The principal closed conformal Killing--Yano tensor of the original metric ${\tb[^{\ind{0}}]{g}}$ ceases to be a closed conformal Killing--Yano tensor for metrics \eqref{eq:igmetrics} with ${i>0}$. It can also be proved that these metrics do not posses a principal closed conformal Killing--Yano tensor which would be compatible with the symmetries discussed below. However, we have not excluded the existence of an unrelated closed conformal Killing--Yano tensor. }

By an appropriate choice of the covariant derivative, namely, ${\tb{\nabla}=\tb[^{\ind{i}}]{\nabla}}$, in the Schouten--Nijenhuis brackets in \eqref{eq:obmutualcom1} and \eqref{eq:obmutualcom2}, we get
\begin{align}
\begin{split}
	\tb[^{\ind{i}}]{\nabla}[^{(a}]\tb[^{\ind{j}}]{k}[^{bc)}] =0\;, \quad \tb[^{\ind{i}}]{\nabla}[^{(a}]\tb[^{\ind{j}}]{l}[^{b)}]=0\;.
\end{split}\label{eq:ikillvecteneq}
\end{align}
Here, the index of ${\tb[^{\ind{i}}]{\nabla}[^{a}]}$ is raised by metric $\tb[^{\ind{i}}]{g}$. We see that quantities  $\tb[^{\ind{j}}]{k}$ and $\tb[^{\ind{k}}]{l}$ are Killing tensors and Killing vectors in the sense of the geometry given by the metric $\tb[^{\ind{i}}]{g}$. In other words, all spaces given by the metrics \eqref{eq:igmetrics} have the same explicit and hidden symmetries as the original off-shell \mbox{Kerr--NUT--(A)dS} geometry.

It suggests that the whole family of geometries \eqref{eq:igmetrics} could share the same properties as the \mbox{Kerr--NUT--(A)dS} spacetimes. It is true for properties related to the classical observables, but, unfortunately, not for the properties related to the operators.

Namely, the geodesic motion is completely integrable for all geometries \eqref{eq:igmetrics}. It is guaranteed by the same set of conditions \eqref{eq:obmutualcom1}--\eqref{eq:obmutualcom3}. The choice of the geometry just specifies which of the conserved quantities in involution plays a role of the Hamiltonian.

However, the commutation of the scalar field operators depends on the choice of the geometry ${\tb[^{\ind{i}}]{g}}$. First, the definitions \eqref{eq:opch} of the field operators themselves depend on the choice of a covariant derivative and one has to decide what is a natural choice. If one uses the same covariant derivative ${\tb{\nabla}=\tb[^{\ind{i}}]{\nabla}}$ in the definitions of all operators ${\op{K}_j}$, the curvature quantities in the anomalous conditions \eqref{eq:opmutualcom4}, \eqref{eq:opmutualcom5} or \eqref{eq:anomric}, are related to this derivative. It turns out that the curvature tensors for each of the metrics ${\tb[^{\ind{i}}]{g}}$ do not share the same structure. Only the curvature of the basic \mbox{Kerr--NUT--(A)dS} metric ${\tb[^{\ind{0}}]{g}}$ is special. In particular, only the Ricci tensor ${\tb[^{\ind{0}}]{\Ric}}$ can be diagonalized in the frame \eqref{eq:e&epsbase} of common eigenvectors of the Killing tensors, all higher Ricci tensors ${\tb[^{\ind{i}}]{\Ric}}$, ${i>0}$, are not diagonal in this frame \cite{Kolar:dipl}. The direct check for particular low dimensions indicates that the anomalous conditions \eqref{eq:opmutualcom5} and \eqref{eq:anomric} are, in general, not satisfied for the higher metrics ${\tb[^{\ind{i}}]{g}}$, ${i>0}$, cf.~\cite{Kolar:dipl}. It means, that although classical variables ${K_i}$ commute between each other and with the Hamiltonian given by ${\tb[^{\ind{i}}]{g}}$, the corresponding operators created using derivative ${\tb[^{\ind{i}}]{\nabla}}$ do not commute.

However, there exists a surprising exception: the anomalous conditions are met for the highest metric ${\tb[^{\ind{n{-}1}}]{g}}$ in an arbitrary dimension \cite{Kolar:unpub}.

\subsection{Electromagnetic field preserving commutation of classical observables}\label{ssc:kerrnutads-elmagob}

A full generalization of the \mbox{Kerr--NUT--(A)dS} spacetimes to the charged case is not known. However, one can study, at least, a weak electromagnetic field on the uncharged background, cf. e.g., \cite{FrolovKrtous:2011,CarigliaEtal:2013a}. The existence of ${2n}$ conserved quantities in involution shows that the geodesic motion in \mbox{Kerr--NUT--(A)dS} spacetimes is a highly special dynamical system: it is completely integrable. It is natural to ask for which test electromagnetic field the charged particle motion remains completely integrable.

We thus want to study conditions of mutual commutativity of classical charged observables ${\ti{q}K_j}$ and ${\ti{q}L_k}$ of the form \eqref{eq:clasobch} with coefficients  $\tb[^{\ind{j}}]{k}$ and $\tb[^{\ind{k}}]{l}$ given by \eqref{eq:klinbase}.
Since classical uncharged observables ${K_j}$ and ${L_k}$ mutually commute, the field $\tb{A}$ must meet conditions \eqref{eq:Aobcond1} and \eqref{eq:Aobcond2}.

Let $\t{A}[_{\mu}]$, $\t{A}[_{\hat{\mu}}]$ denote components\footnote{%
The components $\t{A}[_{\mu}]$ of the vector potential ${\tb{A}}$ are not related to the metric functions ${A_\mu^{(i)}}$.}
of the field $\tb{A}$ with respect to the un-normalized frame ${\epsf{\mu},\,\epsfh{\mu}}$,
\begin{align}
	\tb{A}=\sum_{\mu}\big(\t{A}[_{\mu}]\epsf{\mu}+\t{A}[_{\hat{\mu}}]\epsfh{\mu}\big)\;.
\end{align}
The Killing tensors and vectors \eqref{eq:klinbase} in this frame read
\begin{align}
\begin{split}
	\tb[^{\ind{j}}]{k} &=\sum_{\mu}A_{\mu}^{(j)}\bigg(Q_\mu\epsv{\mu}\epsv{\mu}+\frac{1}{Q_\mu}\epsvh{\mu}\epsvh{\mu}\bigg)\;,\\
	\tb[^{\ind{j}}]{l} &=\sum_\mu A_{\mu}^{(j)}\epsvh{\mu}\;.
\end{split}\label{eq:killtenvecepsilon}
\end{align}

Condition \eqref{eq:Aobcond1} implies that components $\t{A}[_{\mu}]$ and $\t{A}[_{\hat{\mu}}]$ are independent of $\psi_j$,
\begin{equation}\label{eq:Axdep}
    \t{A}[_{\mu}] = \t{A}[_{\mu}](x_1,\dots,x_n)\;,\quad    \t{A}[_{\hat{\mu}}] = \t{A}[_{\hat{\mu}}](x_1,\dots,x_n)\;.
\end{equation}
Now we can turn to condition \eqref{eq:Aobcond2}. Thanks to \eqref{eq:Axdep}, the exterior derivative ${\tb{F}=\tb{\dd}\tb{A}}$ of the vector potential yields
\begin{equation}
\begin{split}
	\tb{F} &= \sum_{\mu,\nu} \t{A}[_{\mu,\nu}]\epsf{\nu}\wedge\epsf{\mu}+ \sum_{\mu,\nu} \t{A}[_{\hat{\mu},\nu}]\epsf{\nu}\wedge\epsfh{\mu} +\sum_{\mu}\t{A}[_{\hat{\mu}}]\tb{\dd}\epsfh{\mu} \\
		&=\sum_{\substack{\mu,\nu\\ \nu\neq\mu}} \Bigg[\t{A}[_{\mu,\nu}]\epsf{\nu}\wedge\epsf{\mu}+\bigg(\t{A}[_{\hat{\mu},\nu}]+\t{A}[_{\hat{\mu}}]\frac{2x_\nu}{x_\nu^2-x_\mu^2}\bigg)\epsf{\nu}\wedge\epsfh{\mu}\Bigg] \\
		 &\feq+\sum_{\nu}\bigg(\t{A}[_{\hat{\nu},\nu}]-\sum_{\substack{\mu\\ \mu\neq\nu}} \t{A}[_{\hat{\mu}}]\frac{2x_\nu}{x_\nu^2-x_\mu^2}\bigg)\epsf{\nu}\wedge\epsfh{\nu}\;,
\end{split} \label{eq:dA}\raisetag{6,5ex}
\end{equation}
where we have substituted
\begin{equation}
\begin{aligned}
	\tb{\dd}\epsf{\mu} &=0\;,\\
	\tb{\dd}\epsfh{\mu} &= \sum_{\substack{\nu\\ \nu\neq\mu}}\frac{2x_\nu}{x_\nu^2-x_\mu^2}\big(\epsf{\nu}\wedge\epsfh{\mu}-\epsf{\nu}\wedge\epsfh{\nu} \big)\;,
\end{aligned}
\end{equation}
cf.~\eqref{eq:e&epsbase} and \eqref{eq:AUidadv}.
Using \eqref{eq:dA} we can rewrite \eqref{eq:Aobcond2} as
\begin{align}
\begin{split}
	0 &= \sum_{\substack{\mu,\nu\\ \nu\neq\mu}} \frac{1}{2}Q_\mu Q_\nu\big(A_{\nu}^{(j)}A_{\mu}^{(k)}-A_{\mu}^{(j)}A_{\nu}^{(k)}\big)\t{A}[_{\mu,\nu}](\epsv{\nu}\epsv{\mu}+\epsv{\mu}\epsv{\nu}) \\
		&\feq +\sum_{\substack{\mu,\nu\\ \nu\neq\mu}} \frac{1}{2}\frac{Q_\nu}{ Q_\mu}\big(A_{\nu}^{(j)}A_{\mu}^{(k)}-A_{\mu}^{(j)}A_{\nu}^{(k)}\big)\\
		 &\feq\feq\times\bigg(\t{A}[_{\hat{\mu},\nu}]+\t{A}[_{\hat{\mu}}]\frac{2x_\nu}{x_\nu^2-x_\mu^2}\bigg)(\epsv{\nu}\epsvh{\mu}+\epsvh{\mu}\epsv{\nu})\;,
\end{split}\raisetag{14ex}
\end{align}
or equivalently
\begin{gather}
	\t{A}[_{[\mu,\nu]}] =0\;, \label{eq:Acondcomp1} \\
	\big(\t{A}[_{\hat{\mu}}](x_\mu^2-x_\nu^2)\big)_{\!,\nu} =0\;. \label{eq:Acondcomp2}
\end{gather}

Equations \eqref{eq:Acondcomp1} and \eqref{eq:Axdep} ensure (locally) the existence of a potential ${\varphi=\varphi(x_1,\dots,x_n)}$
\begin{align}
	\sum_{\mu}\t{A}[_{\mu}]\epsf{\mu}= \tb{\dd}\varphi\;.
\end{align}
Integrating \eqref{eq:Acondcomp2} we find
\begin{align}
	\t{A}[_{\hat{\mu}}]=\frac{f_\mu}{U_\mu}\;, \label{eq:Acomponents}
\end{align}
where ${f_\mu}$ are arbitrary functions of a single variable, ${f_\mu=f_\mu(x_\mu)}$.
Thus, conditions \eqref{eq:Aobcond1} and \eqref{eq:Aobcond2} lead to the field
\begin{equation}
	 \tb{A}= \tb{\dd}\varphi + \sum_{\mu}\frac{f_\mu}{U_\mu}\epsfh{\mu}\;. \label{eq:Afinal}
\end{equation}
Clearly, term $\tb{\dd}\varphi$ is gauge trivial. The Maxwell tensor~${\tb{F}}$ corresponding to \eqref{eq:Afinal} can be obtained from \eqref{eq:dA}, cf. Eqs. \eqref{eq:AUQder} and \eqref{eq:Usum},
\begin{equation}
\begin{split}
	\tb{F} &=\sum_{\nu}\bigg(\t{A}[_{\hat{\nu},\nu}]-\sum_{\substack{\mu\\ \mu\neq\nu}} \t{A}[_{\hat{\mu}}]\frac{2x_\nu}{x_\nu^2-x_\mu^2}\bigg)\epsf{\nu}\wedge\epsfh{\nu} \\	
		&=\sum_{\nu}\bigg(\frac{f'_\nu}{U_\nu}+2x_\nu \sum_{\substack{\mu\\ \mu\neq\nu}} \frac{1}{U_\mu} \frac{f_\mu-f_\nu}{x_\mu^2-x_\nu^2}\bigg)\epsf{\nu}\wedge\epsfh{\nu}\;.
\end{split}\label{eq:elmagten}
\end{equation}

We thus found a general electromagnetic field preserving mutual commutation of conserved quantities and hence the complete integrability of a charged particle motion on \mbox{Kerr--NUT--(A)dS} spacetimes.

Actually, the derivation of the electromagnetic field satisfying \eqref{eq:Aobcond1} and \eqref{eq:Aobcond2} is metric independent [except the assumption of the existence of Killing tensors and Killing vectors \eqref{eq:klinbase}] and therefore the charged particle motion is completely integrable in any of the spaces with metric~$\tb[^{\ind{i}}]{g}$.

The field \eqref{eq:Afinal}, \eqref{eq:elmagten} has been already discussed in a slightly different context. In \cite{Krtous:2007}, it has been derived from the assumptions of the algebraic alignment with background geometry, and in \cite{ChenLu:2008}, its special subcase has appeared in the study of harmonic 2-forms on \mbox{Kerr--NUT--(A)dS} background.

The special case of \eqref{eq:Afinal} is the field, considered in \cite{FrolovKrtous:2011} and \cite{CarigliaEtal:2013b}, of the form
\begin{align}
	\tb{A}=e\,\tb{\xi} \label{eq:killingA}
\end{align}
with ${e}$ being a constant parameter of the field strength. ${\tb{\xi}}$ is the Killing--Yano 1-form given by
\begin{equation}\label{xidef}
  \tb{\xi}= \sum_{\mu}Q_\mu\epsfh{\mu}\;,
\end{equation}
which is actually directly related to Killing vectors as ${\tb{\xi}[_a]= \tb[^{\ind{j}}]{g}[_{ab}]\tb[^{\ind{j}}]{l}[^b]}$ for arbitrary ${j=0,\dots,n-1}$. In this case, components $\t{A}[_{\hat{\mu}}]$ are given by the functions
\begin{align}
f_\mu=e\,X_\mu\;.
\end{align}

Another special case of \eqref{eq:Afinal} is a field satisfying the vacuum Maxwell equations
\begin{equation}\label{eq:MaxwellEq}
  \tb[^{\ind{0}}]{\nabla}[^b]\tb{F}[_{ba}]=0\;.
\end{equation}
The index of the covariant derivative is raised by metric $\tb[^{\ind{0}}]{g}$. Because of the involvement of the metric and its covariant derivative, the source Maxwell equation is metric dependent. It was shown in \cite{Krtous:2007} that the field satisfying \eqref{eq:MaxwellEq} has the form
\begin{align}
	\tb{A}=\sum_{\mu}\frac{e_\mu x_\mu}{U_\mu}\,\epsfh{\mu}\;,
\end{align}
where $e_\mu$ are constants parametrizing the field strength. It corresponds to the choice
\begin{equation}\label{eq:Asrcfree}
	f_\mu=e_\mu x_\mu\;.
\end{equation}

It is not known if the vacuum Maxwell equations can be satisfied in the class of electromagnetic fields \eqref{eq:Afinal} for the higher metrics \eqref{eq:igmetrics}.

\subsection{Separability of the Hamilton--Jacobi equation}\label{ssc:kerrnutads-hj}

It was shown in \cite{FrolovEtal:2007} that not only is geodesic motion in \mbox{Kerr--NUT--(A)dS} spacetimes completely integrable, but also the Hamilton--Jacobi equation can be solved by a separation of variables. Now we demonstrate that this property remains valid also for a charged particle in the test electromagnetic field \eqref{eq:Afinal}.

We consider the field \eqref{eq:Afinal} with the gauge trivial part set to zero, ${\varphi=0}$, but with generic functions ${f_\mu}$. It means that the field does not necessarily satisfy the source-free Maxwell equation. For such a field we can investigate motion of a charged scalar particle simultaneously in all spaces $\tb[^{\ind{i}}]{g}$, because ${\ti{q}K_j}$ and ${\ti{q}L_k}$ are mutually commuting independent conserved quantities in each space $\tb[^{\ind{i}}]{g}$. The corresponding Hamiltonian~${\ti{q}H_i}$ is equal to the observable~${\ti{q}K_i}$.

In what follows, we solve the equations ${\ti{q}K_j=\Xi_j}$ and ${\ti{q}L_j=\Psi_j}$, where $\Xi_j$ and $\Psi_j$ are constants of motion. If we substitute $\tb{p}\equiv\tb{\nabla}S$ in these equations, we obtain
\begin{gather}
	 \big(\tb{\nablai}[_a]S-q\tb{A}[_a]\big)\tb[^{\ind{j}}]{k}[^{ab}]\big(\tb{\nablai}[_b]S-q\tb{A}[_b]\big) =\Xi_j\;, \label{eq:HJchK}\\
	\tb[^{\ind{j}}]{l}[^a]\tb{\nablai}[_a]S =\Psi_j\;. \label{eq:HJchL}
\end{gather}
Equation \eqref{eq:HJchK} for ${j=i}$ is just the Hamilton--Jacobi equation with the time independent Hamiltonian ${\ti{q}H_i}$. By time we mean the world line parameter.

These equations can be simultaneously solved by a separation of variables. We use the following additive-separability ansatz for the Jacobi function
\begin{align}
	S=\sum_\mu S_\mu+\sum_k\Psi_k\psi_k\;, \label{eq:actionansatz}
\end{align}
where $S_\mu$ are functions of a single variable, ${S_\mu=S_\mu(x_\mu)}$. This form automatically guaranties that the conditions \eqref{eq:HJchL} are satisfied.

Substituting \eqref{eq:killtenvecepsilon} and \eqref{eq:Afinal} with ${\varphi=0}$ and \eqref{eq:actionansatz} into \eqref{eq:HJchK} yields
\begin{align}
	 \sum_\mu\frac{A_{\mu}^{(j)}}{U_\mu}\bigg[\big(S^{'}_{\mu}\big)^{\!2}+\frac{1}{X_\mu}\big(\tilde{\Psi}_\mu -q f_\mu\big)^{\!2}\bigg] =\Xi_j\;,
\end{align}
where ${\tilde{\Psi}_\mu\equiv\sum_k\Psi_k\big(\!-\!x_\mu^2\big)^{\!n-1-k}}$. Using \eqref{eq:AUid} we can rewrite this equation as a set of ordinary differential equations for functions $S_\mu$
\begin{align}
	\big(S^{'}_{\mu}\big)^{\!2}=\frac{\tilde{\Xi}_\mu}{X_\mu}-\frac{1}{X_\mu^2}\big(\tilde{\Psi}_\mu -q f_\mu\big)^{\!2}\;, \label{eq:Smudifeq}
\end{align}
where ${\tilde{\Xi}_\mu\equiv\sum_k\Xi_k\big(\!-\!x_\mu^2\big)^{\!n-1-k}}$.

Notice that the differential equation \eqref{eq:Smudifeq} for $S_\mu$ differs from the uncharged case just by the term $q f_\mu$. For a solution of the source-free Maxwell equations the function $f_\mu$ is given by \eqref{eq:Asrcfree} and  Eq. \eqref{eq:Smudifeq} is thus modified only by this term.

Functions $S_\mu$ are fixed by Eq. \eqref{eq:Smudifeq} up to an additive constant. Finally, the Jacobi function $S$ is identical for all metrics $\tb[^{\ind{i}}]{g}$.

\subsection{Electromagnetic field preserving commutation of operators}\label{ssc:kerrnutads-elmagop}

As we discussed above, the definition of the operators \eqref{eq:opch} and its relation depends on the choice of the metric. Therefore, in the following we restrict ourselves to the case of the off-shell \mbox{Kerr--NUT--(A)dS} geometry given by the metric ${\tb{g}=\tb[^{\ind{0}}]{g}}$ and we also set ${\tb{\nabla} = \tb[^{\ind{0}}]{\nabla}}$.\footnote{Actually, everything works also for the metric $\tb[^{\ind{n{-}1}}]{g}$ and the derivative $\tb[^{\ind{n{-}1}}]{\nabla}$, see \cite{Kolar:unpub}. Namely, Eq. \eqref{eq:Aanomcondsatisfied} holds and the Klein-Gordon equation can be solved by a separation of variables.}

The commutation of the charged operators $\ti{q\,}{\op{K}}_i$ and $\ti{q\,}{\op{L}}_j$ with coefficients given by Killing tensors and vectors \eqref{eq:klinbase} leads to conditions \eqref{eq:Aobcond1}, \eqref{eq:Aobcond2} and \eqref{eq:Aopcond3}. Conditions \eqref{eq:Aobcond1} and \eqref{eq:Aobcond2} have been already investigated; they imply that $\tb{A}$ has the form \eqref{eq:Afinal}. It turns out that the last condition \eqref{eq:Aopcond3} is a consequence of the previous. Indeed, since ${\tb{\nablai}[_a]\epsvh{\mu}[a]=0}$ and ${\epsvh{\mu}[a]\,\tb{\dd}[_a]x_\nu=0}$, we have ${\tb{\nablai}[_c]\big(\tb[^{\ind{j}}]{k}[^{cd}]\tb{A}[_d]\big)=\tb{\nablai}[_c]\big(\tb[^{\ind{j}}]{k}[^{cd}]\tb{\nablai}[_d]\,\varphi\big)}$, and the right-hand side of the condition \eqref{eq:Aopcond3} takes the form
\begin{align}
  \tb{\nablai}[_a]\Big(\tb[^{\ind{i}}]{k}[^{ab}]\tb{\nablai}[_b]
  \big(\tb{\nablai}[_c]\big(\tb[^{\ind{j}}]{k}[^{cd}]\tb{A}[_d]\big)\big)\Big) {-}\switch{i}{j}
  =\com{{\op{K}}_i}{{\op{K}}_j}\varphi\,.\label{eq:Aanomcondsatisfied}
\end{align}
It vanishes thanks to the commutation of the uncharged operators ${\op{K}_i}$.

The electromagnetic field \eqref{eq:Afinal} thus preserves the mutual commutation of charged operators \eqref{eq:opch} defined using the \mbox{Kerr--NUT--(A)dS} metric \eqref{eq:kerrnutads} and the corresponding covariant derivative ${\tb{\nabla}}$.

\subsection{Separability of the Klein--Gordon equation}\label{ssc:kerrnutads-kg}

Finally, we generalize the results of \cite{FrolovEtal:2007} and \cite{SergyeyevKrtous:2008} to the charged case. It was shown there that the Klein--Gordon equation and analogous equations for operators $\op{K}_i$ can be solved by a separation of variables.

Charged operators $\ti{q\,}{\op{K}}_i$, $\ti{q\,}{\op{L}}_j$ mutually commute and therefore they have common eigenfunctions $\phi$, which satisfy
\begin{align}
	\ti{q\,}{\op{K}}_j \phi &=\Xi_j \phi\;, \label{eq:KGchK}\\
	\ti{q\,}{\op{L}}_j \phi &=\Psi_j \phi\;, \label{eq:KGchL}
\end{align}
where $\Xi_j$ and $\Psi_j$ are corresponding eigenvalues. Equation \eqref{eq:KGchK} for ${j=0}$ is the Klein--Gordon equation, because ${\ti{q\,}{\op{K}}_0=\ti{q\,}{\Box}}$.

If we chose the trivial gauge ${\varphi=0}$ for the electromagnetic field \eqref{eq:Afinal}, the eigenfunctions can be found in the separated form
\begin{equation}
	\phi = \prod_\mu R_\mu \prod_k\exp{\big(\imag\Psi_k\psi_k\big)}\;,  \label{eq:phiansatz}
\end{equation}
where $R_\mu$ are functions of a single variable, ${R_\mu=R_\mu(x_\mu)}$.

The ansatz \eqref{eq:phiansatz} automatically satisfies \eqref{eq:KGchL}. In order to find functions $R_\mu$ from \eqref{eq:KGchK}, we need to express operator $\ti{q\,}{\op{K}}_j$ in coordinates. Using \eqref{eq:killtenvecepsilon} and \eqref{eq:Afinal} we write each term of this operator [cf. the second equality of $\ti{q\,}{\op{K}}$ in \eqref{eq:opch}] as
\begin{equation}
\begin{split}
-\tb{\nablai}[_a]\tb[^{\ind{j}}]{k}[^{ab}]\tb{\nablai}[_b] &=- \gden^{-\frac12}\tb{\pp}[_a]\gden^{\frac12}\tb[^{\ind{j}}]{k}[^{ab}]\tb{\pp}[_b]\\
	&=-\sum_{\mu} \frac{A_{\mu}^{(j)}}{U_\mu}\Bigg[\frac{\pp}{\pp x_\mu}X_\mu \frac{\pp}{\pp x_\mu} \\
	 &\feq+\frac{1}{X_\mu}\bigg[\sum_k\big(\!-\!x_\mu^2\big)^{\!n-1-k}\frac{\pp}{\pp\psi_k}\bigg]^2\Bigg]\;,\\
2\imag q\tb{A}[_a]\tb[^{\ind{j}}]{k}[^{ab}]\tb{\nablai}[_b] &=2\imag q \sum_\mu \frac{A_{\mu}^{(j)}}{U_\mu}\frac{f_\mu}{X_\mu}\sum_k\big(\!-\!x_\mu^2\big)^{\!n-1-k}\frac{\pp}{\pp\psi_k}\;,\\
\tb{\nablai}[_a]\big(\tb[^{\ind{j}}]{k}[^{ab}]\tb{A}[_b]\big) &={\op{K}}_j\varphi +\gden^{-\frac12}\tb{\pp}[_a]\bigg(\gden^{\frac12}\sum_\mu\frac{f_\mu}{U_\mu}\tb[^{\ind{j}}]{k}[^{ab}]\epsfh{\mu}[b]\bigg)=0\;,\\
q^2\tb{\pow{A}{2}}[_{ab}]\tb[^{\ind{j}}]{k}[^{ab}] &=q^2\sum_\mu \frac{A_{\mu}^{(j)}}{U_\mu} \frac{f_\mu^2}{X_\mu}\;,
\end{split}\raisetag{27ex}
\end{equation}
where $\tb{\pp}$ is the coordinate derivative with respect to the coordinates $x_\mu$, $\psi_k$ and ${\gden=U^2}$ is the determinant of the metric in these coordinates [cf. definition \eqref{eq:functionU} and useful relation \eqref{eq:UdivUmu}]. Putting all this together, operator $\ti{q\,}{\op{K}}_j$ can be written in the form
\begin{equation}\label{eq:opKincoord}
\begin{split}
	\ti{q\,}{\op{K}}_j &=\sum_{\mu} \frac{A_{\mu}^{(j)}}{U_\mu}\Bigg[-\frac{\pp}{\pp x_\mu}X_\mu \frac{\pp}{\pp x_\mu} \\
	 &\feq+\frac{1}{X_\mu}\bigg[-\imag\sum_k\big(\!-\!x_\mu^2\big)^{\!n-1-k}\frac{\pp}{\pp\psi_k}-qf_\mu\bigg]^2\Bigg]\;.
\end{split}\raisetag{13ex}
\end{equation}
Substituting \eqref{eq:phiansatz} and \eqref{eq:opKincoord} into Eq. \eqref{eq:KGchK}, we obtain
\begin{equation}
\begin{split}
	\shiftandleft{2em}\sum_{\mu}\frac{A_{\mu}^{(j)}}{U_\mu}\bigg[-\frac{1}{R_{\mu}}\big(X_\mu R_{\mu}^{'} \big){}^{\! '}+\frac{1}{X_\mu}\big(\tilde{\Psi}_\mu-qf_{\mu}\big)^{\! 2}\bigg] =\Xi_j\;,
\end{split}
\end{equation}
where ${\tilde{\Psi}_\mu\equiv\sum_k\Psi_k\big(\!-\!x_\mu^2\big)^{\!n-1-k}}$. Inverting this relation using \eqref{eq:AUid}, we obtain the set of ordinary differential equations for functions $R_{\mu}$
\begin{equation}\label{eq:Rdifeq}
	\big(X_\mu R_{\mu}^{'} \big){}^{\! '}+\bigg(\tilde{\Xi}_\mu -\frac{1}{X_\mu}\big(\tilde{\Psi}_\mu-qf_{\mu}\big)^{\! 2}\bigg)R_\mu=0\;,
\end{equation}
where ${\tilde{\Xi}_\mu\equiv\sum_k\Xi_k\big(\!-\!x_\mu^2\big)^{\!n-1-k}}$.

We thus proved that the set of the second-order equations \eqref{eq:KGchK} and of the first-order equations \eqref{eq:KGchL}, with operators defined using the standard covariant derivative and the test electromagnetic field \eqref{eq:Afinal}, can be solved by the multiplicative separation of variables \eqref{eq:phiansatz}. Differential equations \eqref{eq:Rdifeq} for functions $R_\mu$ differ from the uncharged case by terms $q f_\mu$. For the electromagnetic field \eqref{eq:Asrcfree}, which solves the source-free Maxwell equations \eqref{eq:MaxwellEq} the character of equations \eqref{eq:Rdifeq} does not change significantly, since the terms ${\tilde{\Psi}_\mu-qf_{\mu}}$ remain polynomial.

\section{Conclusions}\label{sc:concl}

In this paper, we have derived several commutativity conditions of classical observables, as well as field operators, where additional conditions (the so-called anomalous) must hold. We have further generalized classical observables and field operators to the case when a background electromagnetic field is present and found corresponding commutativity conditions. We have concentrated on the physical situation when the manifold is endowed with a metric which defines the Hamiltonian and d'Alembert operator.

We have investigated a fulfillment of derived commutativity conditions in \mbox{Kerr--NUT--(A)dS} spacetimes \eqref{eq:kerrnutads} and spaces \eqref{eq:igmetrics} which share the same set of explicit and hidden symmetries. In particular, we have found the most general electromagnetic field preserving integrability and commutation of operators in \mbox{Kerr--NUT--(A)dS} and related spaces; see \eqref{eq:Afinal}. For such a field we have solved Hamilton--Jacobi and Klein--Gordon equations by the separation of variables. We have thus generalized results of \cite{FrolovKrtous:2011} and \cite{CarigliaEtal:2013b}.

We have introduced the new family of geometries \eqref{eq:igmetrics} with the same explicit and hidden symmetries as the off-shell \mbox{Kerr--NUT--(A)dS} geometry. These related spaces share certain properties with the \mbox{Kerr--NUT--(A)dS} spacetimes. In particular, the motion of a charged particle in electromagnetic field \eqref{eq:Afinal} is completely integrable (classical observables Poisson commute) for all related geometries. Corresponding Hamilton--Jacobi equations can be solved simultaneously by the same Jacobi function. However, an analogous property does not hold in the case of the operators. Only the off-shell \mbox{Kerr--NUT--(A)dS} metric is exceptional because it satisfies necessary anomalous conditions and enables thus the separation of the Klein--Gordon equation.

\section*{Acknowledgments}

The authors would like to thank David Kubiz\v{n}\'{a}k for reading the manuscript and for valuable discussions. This work was supported by the project of excellence of the Czech Science Foundation No.~\mbox{14-37086G}. I.K. has been also supported by the Charles University Grant No.~\mbox{SVV-260211}. 


\appendix
\section*{Appendix: Useful identities}

\subsection{Symmetrization of covariant derivatives}\label{apx:symcov}
Second-order torsion-free covariant derivatives of a scalar field are symmetric ${\tb{{\nablai}}[_a]\tb{{\nablai}}[_b]=\tb{\nablan{2}}[_{ab}]}$. The higher-order derivatives do not have this property, but they can always be split into a symmetric part and a part which contains derivatives of the lower order. In this paper, we have repeatedly employed the following relations:
\begin{align}\label{eq:symkov}
\begin{split}
	\tb{\nablai}[_a]\tb{\nablai}[_b]\tb{\nablai}[_c] &=\tb{\nablan{3}}[_{abc}]-\frac{2}{3}\tb{R}[_{a(b}^{d}_{c)}]\tb{\nablai}[_{d}]\;, \\
\tb{\nablan{2}}[_{ab}]\tb{\nablan{2}}[_{cd}]-\tb{\nablan{2}}[_{cd}]\tb{\nablan{2}}[_{ab}] &=2\big[\tb{R}[_{(c|(a}^e_{b)}]\tb{\nablai}[_{|d)}]\tb{\nablai}[_e]-\tb{R}[_{(a|(c}^e_{d)}]\tb{\nablai}[_{|b)}]\tb{\nablai}[_e]\big] \\
	 &\feq-\big(\tb{\nablai}[_{(a}]\tb{R}[_{b)(c}^e_{d)}]-\tb{\nablai}[_{(c}]\tb{R}[_{d)(a}^e_{b)}]\big)\tb{\nablai}[_e]\; ,
\end{split}\raisetag{13ex}
\end{align}
which can be proved using the Ricci and Bianchi identities.

\subsection{Identities of Schouten--Nijenhuis brackets}\label{apx:snbr}
We list here some identities which were used to obtain \eqref{eq:VLKpoisch} and \eqref{eq:VLKcomch}.
\begin{widetext}
\begin{align}\label{eq:SNidentities}
\begin{split}
  \sn{\tb[^{\ind{i}}]{l}}{\tb{A}\con\tb[^{\ind{j}}]{k}}[a]
     &=\tb{A}[_b]\sn{\tb[^{\ind{i}}]{l}}{\tb[^{\ind{j}}]{k}}[ab]
     +\big(\lie{\tb[^{\ind{i}}]{l}}{\tb{A}[_b]}\big)\tb[^{\ind{j}}]{k}[^{ab}]\;,\\
  \sn{\tb{\pow{A}{2}}[_{ab}]\tb[^{\ind{j}}]{k}[^{ab}]}{\tb[^{\ind{i}}]{l}}
    &=\tb{\pow{A}{2}}[_{ab}]\sn{\tb[^{\ind{j}}]{k}}{\tb[^{\ind{i}}]{l}}[ab]
    -2\big(\lie{\tb[^{\ind{i}}]{l}}{\tb{A}[_a]}\big)\tb[^{\ind{j}}]{k}[^{ab}]\tb{A}[_b]\;,\\
  2\sn{\tb{A}\con\tb[^{\ind{i}}]{k}}{\tb{A}\con\tb[^{\ind{j}}]{k}}[a]
    +\frac{1}{2}\big(\sn{\tb[^{\ind{i}}]{k}}{\tb{\pow{A}{2}}[_{bc}]\tb[^{\ind{j}}]{k}[^{bc}]}[a]-\switch{i}{j}\big)
    &=\frac{3}{2}\tb{\pow{A}{2}}[_{bc}]\sn{\tb[^{\ind{i}}]{k}}{\tb[^{\ind{j}}]{k}}[abc]+4\tb{A}[_b]
    \tb[^{\ind{i}}]{k}[^{c(a}]\tb{F}[_{cd}] \tb[^{\ind{j}}]{k}[^{b)d}]\;,\\
 \sn{\tb{A}\con\tb[^{\ind{i}}]{k}}{\tb[^{\ind{j}}]{k}}[ab]-\switch{i}{j} &=\frac{3}{2}\tb{A}[_c]\sn{\tb[^{\ind{i}}]{k}}{\tb[^{\ind{j}}]{k}}[abc] +2 \tb[^{\ind{i}}]{k}[^{c(a}]\tb{F}[_{cd}] \tb[^{\ind{j}}]{k}[^{b)d}]\;, \\
  \sn{\tb{A}\con\tb[^{\ind{i}}]{k}}{\tb{\pow{A}{2}}[_{bc}]\tb[^{\ind{j}}]{k}[^{bc}]}-\switch{i}{j}
    &=\frac{1}{2}\tb{\pow{A}{3}}[_{abc}]\sn{\tb[^{\ind{i}}]{k}}{\tb[^{\ind{j}}]{k}}[abc] +2\tb{\pow{A}{2}}[_{ab}]\tb[^{\ind{i}}]{k}[^{c(a}]\tb{F}[_{cd}]\tb[^{\ind{j}}]{k}[^{b)d}]\;.
\end{split}\raisetag{21ex}
\end{align}
\end{widetext}
Here, the notation ${\big(\tb{A}\con\tb{k}\big)\tb{}[^a]=\tb{A}[_b]\tb{k}[^{ba}]}$ was employed.

\subsection{Kerr--NUT--(A)dS metric functions}\label{apx:kerrnutads}

We summarize some important identities containing functions $A_{\mu}^{(i)}$, $U_\mu$ which emerge in the definition of the off-shell \mbox{Kerr--NUT--(A)dS} metric \eqref{eq:kerrnutads}.

Functions $A_{\mu}^{(i)}$, $U_\mu$, defined by \eqref{eq:AUfunctions},  satisfy relations
\begin{align}\label{eq:AUid}
\begin{split}
	\sum_\mu A_{\mu}^{(i)} \frac{\big(\!-\!x_\mu^2\big)^{\!n-1-j}}{U_\mu} =\delta_{ij}\;, \\
	\sum_i A_{\mu}^{(i)} \frac{\big(\!-\!x_\nu^2\big)^{\!n-1-i}}{U_\nu} =\delta_{\mu\nu}\;.
\end{split}
\end{align}

Derivatives of $A_{\mu}^{(i)}$, $U_\mu$ can be written as
\begin{align}\label{eq:AUQder}
\begin{split}
	A_{\mu,\nu}^{(i)} &= (1-\delta_{\mu\nu})\frac{2x_\nu}{x_\nu^2-x_\mu^2}\big(A_{\mu}^{(i)}-A_{\nu}^{(i)}\big)\;,\\
	U_{\mu,\nu} &=\delta_{\mu\nu} \sum_{\substack{\rho\\ \rho\neq\mu}} \frac{2x_\mu}{x_\mu^2-x_\rho^2}U_{\mu} +(1-\delta_{\mu\nu})\frac{2x_\nu}{x_\nu^2-x_\mu^2} U_{\mu}\;.
\end{split}\raisetag{13ex}
\end{align}

Combining \eqref{eq:AUid} and \eqref{eq:AUQder} we obtain
\begin{align} \label{eq:AUidadv}
\begin{split}
	\shiftandleft{2em}\sum_i(n-1-i)A_{\mu}^{(i)}\frac{\big(\!-\!x_\nu^2\big)^{\!n-1-i}}{U_\nu} =\delta_{\mu\nu}\sum_{\substack{\rho\\ \rho\neq\mu}}\frac{x_\mu^2}{x_\mu^2-x_\rho^2}\\
	&\feq+(1-\delta_{\mu\nu})\frac{x_\nu^2}{x_\nu^2-x_\mu^2}\;, \\
	\shiftandleft{2em}\sum_i A_{\mu,\alpha}^{(i)}\frac{\big(\!-\!x_\nu^2\big)^{\!n-1-i}}{U_\nu} =(1-\delta_{\mu\alpha})(\delta_{\mu\nu}-\delta_{\alpha\nu})\frac{2x_\alpha}{x_\alpha^2-x_\mu^2}\;.
\end{split}\raisetag{13ex}
\end{align}

Besides these relations, we list here a useful identity
\begin{equation}
 	\frac{1}{U_\nu}\sum_{\substack{\mu\\ \mu\neq\nu}}\frac{1}{x_\mu^2-x_\nu^2}=\sum_{\substack{\mu\\ \mu\neq\nu}}\frac{1}{U_\mu}\frac{1}{x_\nu^2-x_\mu^2}\;. \label{eq:Usum}
\end{equation}

Finally, let us define the function
\begin{equation}
U\equiv\prod_{\mathclap{\substack{\mu,\nu \\ \mu<\nu}}}\big(x_\mu^2-x_\nu^2\big)\;. \label{eq:functionU}
\end{equation}
It appears naturally as a square root of the determinant of the metric \eqref{eq:kerrnutads} in coordinates $x_\mu$, $\psi_k$. It satisfies the relation
\begin{equation}
 	\bigg(\frac{U}{U_\mu}\bigg)_{\!\!\!\!,\mu}=0\;. \label{eq:UdivUmu}
\end{equation}

\vspace*{13ex}



\begin{thebibliography}{33}%
\makeatletter
\providecommand \@ifxundefined [1]{%
 \@ifx{#1\undefined}
}%
\providecommand \@ifnum [1]{%
 \ifnum #1\expandafter \@firstoftwo
 \else \expandafter \@secondoftwo
 \fi
}%
\providecommand \@ifx [1]{%
 \ifx #1\expandafter \@firstoftwo
 \else \expandafter \@secondoftwo
 \fi
}%
\providecommand \natexlab [1]{#1}%
\providecommand \enquote  [1]{``#1''}%
\providecommand \bibnamefont  [1]{#1}%
\providecommand \bibfnamefont [1]{#1}%
\providecommand \citenamefont [1]{#1}%
\providecommand \href@noop [0]{\@secondoftwo}%
\providecommand \href [0]{\begingroup \@sanitize@url \@href}%
\providecommand \@href[1]{\@@startlink{#1}\@@href}%
\providecommand \@@href[1]{\endgroup#1\@@endlink}%
\providecommand \@sanitize@url [0]{\catcode `\\12\catcode `\$12\catcode
  `\&12\catcode `\#12\catcode `\^12\catcode `\_12\catcode `\%12\relax}%
\providecommand \@@startlink[1]{}%
\providecommand \@@endlink[0]{}%
\providecommand \url  [0]{\begingroup\@sanitize@url \@url }%
\providecommand \@url [1]{\endgroup\@href {#1}{\urlprefix }}%
\providecommand \urlprefix  [0]{URL }%
\providecommand \Eprint [0]{\href }%
\providecommand \doibase [0]{http://dx.doi.org/}%
\providecommand \selectlanguage [0]{\@gobble}%
\providecommand \bibinfo  [0]{\@secondoftwo}%
\providecommand \bibfield  [0]{\@secondoftwo}%
\providecommand \translation [1]{[#1]}%
\providecommand \BibitemOpen [0]{}%
\providecommand \bibitemStop [0]{}%
\providecommand \bibitemNoStop [0]{.\EOS\space}%
\providecommand \EOS [0]{\spacefactor3000\relax}%
\providecommand \BibitemShut  [1]{\csname bibitem#1\endcsname}%
\let\auto@bib@innerbib\@empty
\bibitem [{\citenamefont {Emparan}\ and\ \citenamefont
  {Reall}(2008)}]{EmparanReall:2008}%
  \BibitemOpen
  \bibfield  {author} {\bibinfo {author} {\bibfnamefont {R.}~\bibnamefont
  {Emparan}}\ and\ \bibinfo {author} {\bibfnamefont {H.~S.}\ \bibnamefont
  {Reall}},\ }\href {http://www.livingreviews.org/lrr-2008-6} {\bibfield
  {journal} {\bibinfo  {journal} {Living Rev. Rel.}\ }\textbf {\bibinfo
  {volume} {11}},\ \bibinfo {pages} {6} (\bibinfo {year} {2008})},\ \Eprint
  {http://arxiv.org/abs/arXiv: 0801.3471 [hep-th]} {arXiv: 0801.3471 [hep-th]}
  \BibitemShut {NoStop}%
\bibitem [{\citenamefont {Chen}\ \emph {et~al.}(2006)\citenamefont {Chen},
  \citenamefont {L\"u},\ and\ \citenamefont {Pope}}]{ChenLuPope:2006}%
  \BibitemOpen
  \bibfield  {author} {\bibinfo {author} {\bibfnamefont {W.}~\bibnamefont
  {Chen}}, \bibinfo {author} {\bibfnamefont {H.}~\bibnamefont {L\"u}}, \ and\
  \bibinfo {author} {\bibfnamefont {C.~N.}\ \bibnamefont {Pope}},\ }\href@noop
  {} {\bibfield  {journal} {\bibinfo  {journal} {Class. Quantum Grav.}\ }\textbf
  {\bibinfo {volume} {23}},\ \bibinfo {pages} {5323} (\bibinfo {year}
  {2006})},\ \Eprint {http://arxiv.org/abs/arXiv: hep-th/0604125} {arXiv:
  hep-th/0604125} \BibitemShut {NoStop}%
\bibitem [{\citenamefont {Kubiz\v{n}\'ak}\ and\ \citenamefont
  {Frolov}(2007)}]{KubiznakFrolov:2007}%
  \BibitemOpen
  \bibfield  {author} {\bibinfo {author} {\bibfnamefont {D.}~\bibnamefont
  {Kubiz\v{n}\'ak}}\ and\ \bibinfo {author} {\bibfnamefont {V.~P.}\
  \bibnamefont {Frolov}},\ }\href@noop {} {\bibfield  {journal} {\bibinfo
  {journal} {Class. Quantum Grav.}\ }\textbf {\bibinfo {volume} {24}},\
  \bibinfo {pages} {F1} (\bibinfo {year} {2007})},\ \Eprint
  {http://arxiv.org/abs/arXiv: gr-qc/0610144} {arXiv: gr-qc/0610144}
  \BibitemShut {NoStop}%
\bibitem [{\citenamefont {Krtou\v{s}}\ \emph
  {et~al.}(2007{\natexlab{a}})\citenamefont {Krtou\v{s}}, \citenamefont
  {Kubiz\v{n}\'ak}, \citenamefont {Page},\ and\ \citenamefont
  {Frolov}}]{KrtousEtal:2007a}%
  \BibitemOpen
  \bibfield  {author} {\bibinfo {author} {\bibfnamefont {P.}~\bibnamefont
  {Krtou\v{s}}}, \bibinfo {author} {\bibfnamefont {D.}~\bibnamefont
  {Kubiz\v{n}\'ak}}, \bibinfo {author} {\bibfnamefont {D.~N.}\ \bibnamefont
  {Page}}, \ and\ \bibinfo {author} {\bibfnamefont {V.~P.}\ \bibnamefont
  {Frolov}},\ }\href@noop {} {\bibfield  {journal} {\bibinfo  {journal} {J.
  High Energy Phys.}\ }\textbf {\bibinfo {volume} {0702}},\ \bibinfo {pages}
  {004} (\bibinfo {year} {2007}{\natexlab{a}})},\ \Eprint
  {http://arxiv.org/abs/arXiv: hep-th/0612029} {arXiv: hep-th/0612029}
  \BibitemShut {NoStop}%
\bibitem [{\citenamefont {Page}\ \emph {et~al.}(2007)\citenamefont {Page},
  \citenamefont {Kubiz\v{n}\'ak}, \citenamefont {Vasudevan},\ and\
  \citenamefont {Krtou\v{s}}}]{PageEtal:2007}%
  \BibitemOpen
  \bibfield  {author} {\bibinfo {author} {\bibfnamefont {D.~N.}\ \bibnamefont
  {Page}}, \bibinfo {author} {\bibfnamefont {D.}~\bibnamefont
  {Kubiz\v{n}\'ak}}, \bibinfo {author} {\bibfnamefont {M.}~\bibnamefont
  {Vasudevan}}, \ and\ \bibinfo {author} {\bibfnamefont {P.}~\bibnamefont
  {Krtou\v{s}}},\ }\href@noop {} {\bibfield  {journal} {\bibinfo  {journal}
  {Phys. Rev. Lett.}\ }\textbf {\bibinfo {volume} {98}},\ \bibinfo {pages}
  {061102} (\bibinfo {year} {2007})},\ \Eprint {http://arxiv.org/abs/arXiv:
  hep-th/0611083} {arXiv: hep-th/0611083} \BibitemShut {NoStop}%
\bibitem [{\citenamefont {Frolov}\ \emph {et~al.}(2007)\citenamefont {Frolov},
  \citenamefont {Krtou\v{s}},\ and\ \citenamefont
  {Kubiz\v{n}\'ak}}]{FrolovEtal:2007}%
  \BibitemOpen
  \bibfield  {author} {\bibinfo {author} {\bibfnamefont {V.~P.}\ \bibnamefont
  {Frolov}}, \bibinfo {author} {\bibfnamefont {P.}~\bibnamefont {Krtou\v{s}}},
  \ and\ \bibinfo {author} {\bibfnamefont {D.}~\bibnamefont {Kubiz\v{n}\'ak}},\
  }\href@noop {} {\bibfield  {journal} {\bibinfo  {journal} {J. High Energy
  Phys.}\ }\textbf {\bibinfo {volume} {0702}},\ \bibinfo {pages} {005}
  (\bibinfo {year} {2007})},\ \Eprint {http://arxiv.org/abs/arXiv:
  hep-th/0611245} {arXiv: hep-th/0611245} \BibitemShut {NoStop}%
\bibitem [{\citenamefont {Sergyeyev}\ and\ \citenamefont
  {Krtou\v{s}}(2008)}]{SergyeyevKrtous:2008}%
  \BibitemOpen
  \bibfield  {author} {\bibinfo {author} {\bibfnamefont {A.}~\bibnamefont
  {Sergyeyev}}\ and\ \bibinfo {author} {\bibfnamefont {P.}~\bibnamefont
  {Krtou\v{s}}},\ }\href@noop {} {\bibfield  {journal} {\bibinfo  {journal}
  {Phys. Rev. D}\ }\textbf {\bibinfo {volume} {77}},\ \bibinfo {pages} {044033}
  (\bibinfo {year} {2008})},\ \Eprint {http://arxiv.org/abs/arXiv: 0711.4623
  [hep-th]} {arXiv: 0711.4623 [hep-th]} \BibitemShut {NoStop}%
\bibitem [{\citenamefont {Cariglia}\ \emph {et~al.}(2011)\citenamefont
  {Cariglia}, \citenamefont {Krtou\v{s}},\ and\ \citenamefont
  {Kubiz\v{n}\'ak}}]{CarigliaEtal:2011b}%
  \BibitemOpen
  \bibfield  {author} {\bibinfo {author} {\bibfnamefont {M.}~\bibnamefont
  {Cariglia}}, \bibinfo {author} {\bibfnamefont {P.}~\bibnamefont
  {Krtou\v{s}}}, \ and\ \bibinfo {author} {\bibfnamefont {D.}~\bibnamefont
  {Kubiz\v{n}\'ak}},\ }\href {\doibase 10.1103/PhysRevD.84.024008} {\bibfield
  {journal} {\bibinfo  {journal} {Phys. Rev. D}\ }\textbf {\bibinfo {volume}
  {84}},\ \bibinfo {pages} {024008} (\bibinfo {year} {2011})},\ \Eprint
  {http://arxiv.org/abs/arXiv: 1104.4123 [hep-th]} {arXiv: 1104.4123 [hep-th]}
  \BibitemShut {NoStop}%
\bibitem [{\citenamefont {Houri}\ \emph {et~al.}(2008)\citenamefont {Houri},
  \citenamefont {Oota},\ and\ \citenamefont {Yasui}}]{HouriEtal:2008}%
  \BibitemOpen
  \bibfield  {author} {\bibinfo {author} {\bibfnamefont {T.}~\bibnamefont
  {Houri}}, \bibinfo {author} {\bibfnamefont {T.}~\bibnamefont {Oota}}, \ and\
  \bibinfo {author} {\bibfnamefont {Y.}~\bibnamefont {Yasui}},\ }\href@noop {}
  {\bibfield  {journal} {\bibinfo  {journal} {J. Phys.}\ }\textbf {\bibinfo
  {volume} {A41}},\ \bibinfo {pages} {025204} (\bibinfo {year} {2008})},\
  \Eprint {http://arxiv.org/abs/arXiv: 0707.4039 [hep-th]} {arXiv: 0707.4039
  [hep-th]} \BibitemShut {NoStop}%
\bibitem [{\citenamefont {Houri}\ \emph {et~al.}(2007)\citenamefont {Houri},
  \citenamefont {Oota},\ and\ \citenamefont {Yasui}}]{HouriEtal:2007}%
  \BibitemOpen
  \bibfield  {author} {\bibinfo {author} {\bibfnamefont {T.}~\bibnamefont
  {Houri}}, \bibinfo {author} {\bibfnamefont {T.}~\bibnamefont {Oota}}, \ and\
  \bibinfo {author} {\bibfnamefont {Y.}~\bibnamefont {Yasui}},\ }\href@noop {}
  {\bibfield  {journal} {\bibinfo  {journal} {Phys. Lett.}\ }\textbf {\bibinfo
  {volume} {B656}},\ \bibinfo {pages} {214} (\bibinfo {year} {2007})},\ \Eprint
  {http://arxiv.org/abs/arXiv: 0708.1368 [hep-th]} {arXiv: 0708.1368 [hep-th]}
  \BibitemShut {NoStop}%
\bibitem [{\citenamefont {Krtou{\v s}}\ \emph {et~al.}(2008)\citenamefont
  {Krtou{\v s}}, \citenamefont {Frolov},\ and\ \citenamefont
  {Kubiz\v{n}\'ak}}]{KrtousFrolovKubiznak:2008}%
  \BibitemOpen
  \bibfield  {author} {\bibinfo {author} {\bibfnamefont {P.}~\bibnamefont
  {Krtou{\v s}}}, \bibinfo {author} {\bibfnamefont {V.~P.}\ \bibnamefont
  {Frolov}}, \ and\ \bibinfo {author} {\bibfnamefont {D.}~\bibnamefont
  {Kubiz\v{n}\'ak}},\ }\href@noop {} {\bibfield  {journal} {\bibinfo  {journal}
  {Phys. Rev. D}\ }\textbf {\bibinfo {volume} {78}},\ \bibinfo {pages} {064022}
  (\bibinfo {year} {2008})},\ \Eprint {http://arxiv.org/abs/arXiv: 0804.4705
  [hep-th]} {arXiv: 0804.4705 [hep-th]} \BibitemShut {NoStop}%
\bibitem [{\citenamefont {Aliev}\ and\ \citenamefont
  {Frolov}(2004)}]{AlievFrolov:2004}%
  \BibitemOpen
  \bibfield  {author} {\bibinfo {author} {\bibfnamefont {A.~N.}\ \bibnamefont
  {Aliev}}\ and\ \bibinfo {author} {\bibfnamefont {V.~P.}\ \bibnamefont
  {Frolov}},\ }\href@noop {} {\bibfield  {journal} {\bibinfo  {journal} {Phys.
  Rev. D}\ }\textbf {\bibinfo {volume} {69}},\ \bibinfo {pages} {084022}
  (\bibinfo {year} {2004})},\ \Eprint {http://arxiv.org/abs/arXiv:
  hep-th/0401095} {arXiv: hep-th/0401095} \BibitemShut {NoStop}%
\bibitem [{\citenamefont {Aliev}(2006)}]{Aliev:2006a}%
  \BibitemOpen
  \bibfield  {author} {\bibinfo {author} {\bibfnamefont {A.~N.}\ \bibnamefont
  {Aliev}},\ }\href@noop {} {\bibfield  {journal} {\bibinfo  {journal} {Mod.
  Phys. Lett.}\ }\textbf {\bibinfo {volume} {A21}},\ \bibinfo {pages} {751}
  (\bibinfo {year} {2006})},\ \Eprint {http://arxiv.org/abs/arXiv:
  gr-qc/0505003} {arXiv: gr-qc/0505003} \BibitemShut {NoStop}%
\bibitem [{\citenamefont {Aliev}(2006)}]{Aliev:2006}%
  \BibitemOpen
  \bibfield  {author} {\bibinfo {author} {\bibfnamefont {A.~N.}\ \bibnamefont
  {Aliev}},\ }\href@noop {} {\bibfield  {journal} {\bibinfo  {journal} {Phys.
  Rev. D}\ }\textbf {\bibinfo {volume} {74}},\ \bibinfo {pages} {024011}
  (\bibinfo {year} {2006})},\ \Eprint {http://arxiv.org/abs/arXiv:
  hep-th/0604207} {arXiv: hep-th/0604207} \BibitemShut {NoStop}%
\bibitem [{\citenamefont {Aliev}(2007)}]{Aliev:2007}%
  \BibitemOpen
  \bibfield  {author} {\bibinfo {author} {\bibfnamefont {A.~N.}\ \bibnamefont
  {Aliev}},\ }\href@noop {} {\bibfield  {journal} {\bibinfo  {journal} {Phys.
  Rev. D}\ }\textbf {\bibinfo {volume} {75}},\ \bibinfo {pages} {084041}
  (\bibinfo {year} {2007})},\ \Eprint {http://arxiv.org/abs/arXiv:
  hep-th/0702129} {arXiv: hep-th/0702129} \BibitemShut {NoStop}%
\bibitem [{\citenamefont {Krtou{\v s}}(2007)}]{Krtous:2007}%
  \BibitemOpen
  \bibfield  {author} {\bibinfo {author} {\bibfnamefont {P.}~\bibnamefont
  {Krtou{\v s}}},\ }\href@noop {} {\bibfield  {journal} {\bibinfo  {journal}
  {Phys. Rev. D}\ }\textbf {\bibinfo {volume} {76}},\ \bibinfo {pages} {084035}
  (\bibinfo {year} {2007})},\ \Eprint {http://arxiv.org/abs/arXiv: 0707.0002
  [hep-th]} {arXiv: 0707.0002 [hep-th]} \BibitemShut {NoStop}%
\bibitem [{\citenamefont {Frolov}\ and\ \citenamefont
  {Shoom}(2010)}]{FrolovShoom:2010}%
  \BibitemOpen
  \bibfield  {author} {\bibinfo {author} {\bibfnamefont {V.~P.}\ \bibnamefont
  {Frolov}}\ and\ \bibinfo {author} {\bibfnamefont {A.~A.}\ \bibnamefont
  {Shoom}},\ }\href {\doibase 10.1103/PhysRevD.82.084034} {\bibfield  {journal}
  {\bibinfo  {journal} {Phys. Rev. D}\ }\textbf {\bibinfo {volume} {82}},\
  \bibinfo {pages} {084034} (\bibinfo {year} {2010})},\ \Eprint
  {http://arxiv.org/abs/arXiv: 1008.2985 [gr-qc]} {arXiv: 1008.2985 [gr-qc]}
  \BibitemShut {NoStop}%
\bibitem [{\citenamefont {Aliev}\ and\ \citenamefont
  {Gal'tsov}(1989)}]{AlievGaltsov:1989b}%
  \BibitemOpen
  \bibfield  {author} {\bibinfo {author} {\bibfnamefont {A.~N.}\ \bibnamefont
  {Aliev}}\ and\ \bibinfo {author} {\bibfnamefont {D.~V.}\ \bibnamefont
  {Gal'tsov}},\ }\href@noop {} {\bibfield  {journal} {\bibinfo  {journal} {Sov.
  Phys. Usp.}\ }\textbf {\bibinfo {volume} {32}},\ \bibinfo {pages} {75}
  (\bibinfo {year} {1989})}\BibitemShut {NoStop}%
\bibitem [{\citenamefont {Frolov}\ and\ \citenamefont {Krtou{\v
  s}}(2011)}]{FrolovKrtous:2011}%
  \BibitemOpen
  \bibfield  {author} {\bibinfo {author} {\bibfnamefont {V.~P.}\ \bibnamefont
  {Frolov}}\ and\ \bibinfo {author} {\bibfnamefont {P.}~\bibnamefont {Krtou{\v
  s}}},\ }\href {\doibase 10.1103/PhysRevD.83.024016} {\bibfield  {journal}
  {\bibinfo  {journal} {Phys. Rev. D}\ }\textbf {\bibinfo {volume} {83}},\
  \bibinfo {pages} {024016} (\bibinfo {year} {2011})},\ \Eprint
  {http://arxiv.org/abs/arXiv: 1010.2266 [hep-th]} {arXiv: 1010.2266 [hep-th]}
  \BibitemShut {NoStop}%
\bibitem [{\citenamefont {Cariglia}\ \emph
  {et~al.}(2013{\natexlab{a}})\citenamefont {Cariglia}, \citenamefont {Frolov},
  \citenamefont {Krtou\v{s}},\ and\ \citenamefont
  {Kubiz\v{n}\'ak}}]{CarigliaEtal:2013b}%
  \BibitemOpen
  \bibfield  {author} {\bibinfo {author} {\bibfnamefont {M.}~\bibnamefont
  {Cariglia}}, \bibinfo {author} {\bibfnamefont {V.~P.}\ \bibnamefont
  {Frolov}}, \bibinfo {author} {\bibfnamefont {P.}~\bibnamefont {Krtou\v{s}}},
  \ and\ \bibinfo {author} {\bibfnamefont {D.}~\bibnamefont {Kubiz\v{n}\'ak}},\
  }\href {\doibase 10.1103/PhysRevD.87.064003} {\bibfield  {journal} {\bibinfo
  {journal} {Phys. Rev. D}\ }\textbf {\bibinfo {volume} {87}},\ \bibinfo
  {pages} {064003} (\bibinfo {year} {2013}{\natexlab{a}})},\ \Eprint
  {http://arxiv.org/abs/arXiv: 1211.4631 [gr-qc]} {arXiv: 1211.4631 [gr-qc]}
  \BibitemShut {NoStop}%
\bibitem [{\citenamefont {Schouten}(1940)}]{Schouten:1940}%
  \BibitemOpen
  \bibfield  {author} {\bibinfo {author} {\bibfnamefont {J.~A.}\ \bibnamefont
  {Schouten}},\ }\href@noop {} {\bibfield  {journal} {\bibinfo  {journal}
  {Indag. Math.}\ }\textbf {\bibinfo {volume} {2}},\ \bibinfo {pages} {449}
  (\bibinfo {year} {1940})}\BibitemShut {NoStop}%
\bibitem [{\citenamefont {Schouten}(1954)}]{Schouten:1954}%
  \BibitemOpen
  \bibfield  {author} {\bibinfo {author} {\bibfnamefont {J.~A.}\ \bibnamefont
  {Schouten}},\ }\href@noop {} {\bibfield  {journal} {\bibinfo  {journal}
  {Convegno Int. Geom. Diff.}\ ,\ \bibinfo {pages} {1}} (\bibinfo {year}
  {1954})}\BibitemShut {NoStop}%
\bibitem [{\citenamefont {Nijenhuis}(1955)}]{Nijenhuis:1955}%
  \BibitemOpen
  \bibfield  {author} {\bibinfo {author} {\bibfnamefont {A.}~\bibnamefont
  {Nijenhuis}},\ }\href@noop {} {\bibfield  {journal} {\bibinfo  {journal}
  {Indag. Math.}\ }\textbf {\bibinfo {volume} {17}},\ \bibinfo {pages} {390}
  (\bibinfo {year} {1955})}\BibitemShut {NoStop}%
\bibitem [{\citenamefont {Carter}(1977)}]{Carter:1977}%
  \BibitemOpen
  \bibfield  {author} {\bibinfo {author} {\bibfnamefont {B.}~\bibnamefont
  {Carter}},\ }\href {\doibase 10.1103/PhysRevD.16.3395} {\bibfield  {journal}
  {\bibinfo  {journal} {Phys. Rev. D}\ }\textbf {\bibinfo {volume} {16}},\
  \bibinfo {pages} {3395} (\bibinfo {year} {1977})}\BibitemShut {NoStop}%
\bibitem [{\citenamefont {Arnold}(1989)}]{Arnold:book}%
  \BibitemOpen
  \bibfield  {author} {\bibinfo {author} {\bibfnamefont {V.~I.}\ \bibnamefont
  {Arnol'd}},\ }\href@noop {} {\emph {\bibinfo {title} {Mathematical Methods of Classical Mechanics}}}\ (\bibinfo  {publisher} {Springer},\ \bibinfo
  {address} {New York},\ \bibinfo {year} {1989})\BibitemShut {NoStop}%
\bibitem [{\citenamefont {Cariglia}\ \emph
  {et~al.}(2013{\natexlab{b}})\citenamefont {Cariglia}, \citenamefont {Frolov},
  \citenamefont {Krtou\v{s}},\ and\ \citenamefont
  {Kubiz\v{n}\'ak}}]{CarigliaEtal:2013a}%
  \BibitemOpen
  \bibfield  {author} {\bibinfo {author} {\bibfnamefont {M.}~\bibnamefont
  {Cariglia}}, \bibinfo {author} {\bibfnamefont {V.~P.}\ \bibnamefont
  {Frolov}}, \bibinfo {author} {\bibfnamefont {P.}~\bibnamefont {Krtou\v{s}}},
  \ and\ \bibinfo {author} {\bibfnamefont {D.}~\bibnamefont {Kubiz\v{n}\'ak}},\
  }\href {\doibase 10.1103/PhysRevD.87.024002} {\bibfield  {journal} {\bibinfo
  {journal} {Phys. Rev. D}\ }\textbf {\bibinfo {volume} {87}},\ \bibinfo
  {pages} {024002} (\bibinfo {year} {2013}{\natexlab{b}})},\ \Eprint
  {http://arxiv.org/abs/arXiv: 1210.3079 [math-ph]} {arXiv: 1210.3079
  [math-ph]} \BibitemShut {NoStop}%
\bibitem [{\citenamefont {Cariglia}(2004)}]{Cariglia:2004}%
  \BibitemOpen
  \bibfield  {author} {\bibinfo {author} {\bibfnamefont {M.}~\bibnamefont
  {Cariglia}},\ }\href@noop {} {\bibfield  {journal} {\bibinfo  {journal}
  {Class. Quantum Grav.}\ }\textbf {\bibinfo {volume} {21}},\ \bibinfo {pages}
  {1051} (\bibinfo {year} {2004})},\ \Eprint {http://arxiv.org/abs/arXiv:
  hep-th/0305153} {arXiv: hep-th/0305153} \BibitemShut {NoStop}%
\bibitem [{\citenamefont {Krtou{\v s}}(2015)}]{Krtous:inprep2015}%
  \BibitemOpen
  \bibfield  {author} {\bibinfo {author} {\bibfnamefont {P.}~\bibnamefont
  {Krtou{\v s}}},\ }\href@noop {} {\  (\bibinfo {year} {2015})},\ \bibinfo
  {note} {to be published}\BibitemShut {NoStop}%
\bibitem [{\citenamefont {Krtou\v{s}}\ \emph
  {et~al.}(2007{\natexlab{b}})\citenamefont {Krtou\v{s}}, \citenamefont
  {Kubiz\v{n}\'ak}, \citenamefont {Page},\ and\ \citenamefont
  {Vasudevan}}]{KrtousEtal:2007b}%
  \BibitemOpen
  \bibfield  {author} {\bibinfo {author} {\bibfnamefont {P.}~\bibnamefont
  {Krtou\v{s}}}, \bibinfo {author} {\bibfnamefont {D.}~\bibnamefont
  {Kubiz\v{n}\'ak}}, \bibinfo {author} {\bibfnamefont {D.~N.}\ \bibnamefont
  {Page}}, \ and\ \bibinfo {author} {\bibfnamefont {M.}~\bibnamefont
  {Vasudevan}},\ }\href@noop {} {\bibfield  {journal} {\bibinfo  {journal}
  {Phys. Rev. D}\ }\textbf {\bibinfo {volume} {76}},\ \bibinfo {pages} {084034}
  (\bibinfo {year} {2007}{\natexlab{b}})},\ \Eprint
  {http://arxiv.org/abs/arXiv: 0707.0001 [hep-th]} {arXiv: 0707.0001 [hep-th]}
  \BibitemShut {NoStop}%
\bibitem [{\citenamefont {Hamamoto}\ \emph {et~al.}(2007)\citenamefont
  {Hamamoto}, \citenamefont {Houri}, \citenamefont {Oota},\ and\ \citenamefont
  {Yasui}}]{HamamotoEtal:2007}%
  \BibitemOpen
  \bibfield  {author} {\bibinfo {author} {\bibfnamefont {N.}~\bibnamefont
  {Hamamoto}}, \bibinfo {author} {\bibfnamefont {T.}~\bibnamefont {Houri}},
  \bibinfo {author} {\bibfnamefont {T.}~\bibnamefont {Oota}}, \ and\ \bibinfo
  {author} {\bibfnamefont {Y.}~\bibnamefont {Yasui}},\ }\href@noop {}
  {\bibfield  {journal} {\bibinfo  {journal} {J. Phys.}\ }\textbf {\bibinfo
  {volume} {A40}},\ \bibinfo {pages} {F177} (\bibinfo {year} {2007})},\ \Eprint
  {http://arxiv.org/abs/arXiv: hep-th/0611285} {arXiv: hep-th/0611285}
  \BibitemShut {NoStop}%
\bibitem [{\citenamefont {Kol\'a\v{r}}(2014)}]{Kolar:dipl}%
  \BibitemOpen
  \bibfield  {author} {\bibinfo {author} {\bibfnamefont {I.}~\bibnamefont
  {Kol\'a\v{r}}},\ }\href@noop {} {Master's
  thesis},\ \bibinfo  {school} {Charles University}, \bibinfo {address}
  {Prague, Czech Republic} (\bibinfo {year} {2014}),\ \bibinfo {note} {in
  Czech}\BibitemShut {NoStop}%
\bibitem [{\citenamefont {Kol\'a\v{r}}()}]{Kolar:unpub}%
  \BibitemOpen
  \bibfield  {author} {\bibinfo {author} {\bibfnamefont {I.}~\bibnamefont
  {Kol\'a\v{r}}},\ }\href@noop {} {\ }\bibinfo {note} {to be published}\BibitemShut
  {NoStop}%
\bibitem [{\citenamefont {Chen}\ and\ \citenamefont
  {L\"u}(2008)}]{ChenLu:2008}%
  \BibitemOpen
  \bibfield  {author} {\bibinfo {author} {\bibfnamefont {W.}~\bibnamefont
  {Chen}}\ and\ \bibinfo {author} {\bibfnamefont {H.}~\bibnamefont {L\"u}},\
  }\href@noop {} {\bibfield  {journal} {\bibinfo  {journal} {Phys. Lett.}\
  }\textbf {\bibinfo {volume} {B658}},\ \bibinfo {pages} {158} (\bibinfo {year}
  {2008})},\ \Eprint {http://arxiv.org/abs/arXiv: 0705.4471 [hep-th]} {arXiv:
  0705.4471 [hep-th]} \BibitemShut {NoStop}%
\end{thebibliography}

%

\end{document}